\documentclass[
  a4paper,
  onecolumn,
  11pt,
  accepted=2026-05-19
]{quantumarticle}

\pdfoutput=1

\usepackage[T1]{fontenc}
\usepackage[utf8]{inputenc}

\usepackage{graphicx}
\usepackage{mathtools}
\usepackage{amsmath,amssymb}
\usepackage{bm}
\usepackage{yhmath}
\usepackage{mathrsfs}
\usepackage{dsfont}
\usepackage{bbold}

\usepackage{dcolumn}
\usepackage[export]{adjustbox}

\usepackage{xcolor}
\usepackage{cancel}
\usepackage{csquotes}
\usepackage{gensymb}
\usepackage{braket}
\usepackage{orcidlink}

\usepackage[numbers,sort&compress]{natbib}

\usepackage{hyperref}
\hypersetup{
    colorlinks=true,
    linktocpage=true,
    citecolor=blue,
    linkcolor=blue,
    urlcolor=blue,
    pdfdisplaydoctitle=true,
    pdfpagemode=UseOutlines,
    bookmarksnumbered=true
}

\renewcommand{\min}{\mathrm{Min}}

\definecolor{cbl}{rgb}{0,0,1}
\definecolor{crd}{rgb}{1,0,0}
\definecolor{Blue}{rgb}{0.0, 0.0, 0.5}

\newcommand{\be}{\begin{equation}}
\newcommand{\ee}{\end{equation}}
\newcommand{\bsplit}{\begin{split}}
\newcommand{\esplit}{\end{split}}

\graphicspath{{Figures/PNG/}{Figures/PDF/}{Figures/EPS/}{Figures/TEX/}{Figures/}}

\begin{document}

\newcommand{\titleinfo}{
{Long-time Freeness in the Kicked Top }
}

\title{\titleinfo}

\author{Elisa Vallini}
\affiliation{Institut f\"ur Theoretische Physik, Universit\"at zu K\"oln, Z\"ulpicher Straße 77, 50937 K\"oln, Germany}

\author{Silvia Pappalardi}
\affiliation{Institut f\"ur Theoretische Physik, Universit\"at zu K\"oln, Z\"ulpicher Straße 77, 50937 K\"oln, Germany}

\begin{abstract}
Recent work highlighted the importance of higher-order correlations in quantum dynamics for a deeper understanding of quantum chaos and thermalization.
The full Eigenstate Thermalization Hypothesis, the framework encompassing correlations, can be formalized using the language of Free Probability theory. In this context, chaotic dynamics at long times are proposed to lead to free independence or ``freeness'' of observables.
In this work, we investigate these issues in a paradigmatic semiclassical model -- the kicked top -- which exhibits a transition from integrability to chaos. Despite its simplicity, we identify several non-trivial features. By numerically studying $2n$-point out-of-time-order correlators, we show that in the fully chaotic regime, long-time freeness is reached exponentially fast. 
These considerations lead us to introduce a \emph{large deviation theory for freeness} that enables us to define and analyze the associated time scale. The numerical results confirm the existence of a hierarchy of different time scales, indicating a multifractal approach to freeness in this model. 
Our findings provide novel insights into the long-time behaviour of chaotic dynamics and may have broader implications for the study of many-body quantum dynamics.
\end{abstract}

\maketitle 

\tableofcontents

\section{Introduction}

The study of quantum chaos and the mechanisms by which isolated quantum systems approach thermal equilibrium has fundamentally reshaped our understanding of quantum theory. Recently, attention has shifted toward more refined probes of chaos, with a particular focus on the role of \emph{higher-order correlations} in quantum dynamics. This shift is motivated by developments at the intersection of various fields. 
In quantum information theory, higher-order moments play a crucial role in $n$-designs, ensembles of unitaries that approximate the statistical properties of uniformly random unitaries up to the $n$-th moment \cite{delsarte1991spherical, dankert2005efficient, dankert2009exact}. This framework has inspired the study of thermalization through higher moments of state ensembles, such as those generated via time evolution or projected ensembles \cite{cotler2023emergent, Choi_2023, ho2022exact, claeys2022emergent, ippoliti2022solvable, lucas2022generalized, bhore2023deep, mcginley2022shadow, pilatowsky2023complete, mark2024maximum}. New insights into the correlation properties of time-evolved states have also been achieved recently by studies on the statistics of energy eigenstates \cite{dymarsky2016subsystem,huang2021universal,lu2019renyi,murthy2019structure,shi2023local,hahn2023statistical, jindal2024generalized}. \\
Concurrently, multi-time correlation functions of physical observables $\hat A$ have gained prominence for their importance in higher-order hydrodynamics beyond the linear response regime \cite{doyon2020fluctuations, myers2020transport, fava2021hydrodynamic, delacretaz2024nonlinear, kawamoto2024strategy} or for the study of quantum information scrambling, quantified by out-of-time-order correlators (OTOCs) $\braket{\hat A(t) \hat A \hat A(t) \hat A}$ \cite{maldacena2016bound, hosur2016chaos, xu2022scrambling, garcia2022out}.
These correlators, which involve unconventional time orderings between times, are known to provide insights into quantum chaos that go beyond standard two-time correlators. For example, in the case of underlying chaotic semi-classical dynamics, the OTOCs encode the quantum Lyapunov exponent at short intermediate times \cite{larkin1969quasiclassical, KitaTalk}. To address properties among $2n$ times, the so-called $2n$-OTOCs 
\begin{equation}
    \label{2notoc}
    \langle (\hat A(t) \hat A )^n \rangle \ ,
\end{equation}
have been intensively studied \cite{roberts2017chaos, cotler2017chaos, tsuji2018bound, cotler2020spectral, leone2021quantum}. These account for $n$-designs information \cite{roberts2017chaos}, and may encode the generalized quantum Lyapunov exponents \cite{Pappalardi2023quantum, trunin2023quantum, trunin2023refined}.

The established framework for understanding quantum dynamics of observables $\hat A$ is the Eigenstate Thermalization Hypothesis (ETH) \cite{srednicki1994chaos, srednicki1999approach, dalessio2016from}. ETH assumes that matrix elements of $\hat A$ in the energy eigenbasis look like pseudorandom matrices with smooth statistical properties. To describe multi-time correlation functions such as in Eq.~\eqref{2notoc}, a complete version of ETH has been introduced, named \emph{full ETH}, which encompasses correlations among the matrix elements \cite{foini2019eigenstate}. The significance of these correlations has attracted substantial interest across fields \cite{prosen1999, sonner2017eigenstate, foini2019eigenstate2, chan2019eigenstate, murthy2019bounds, richter2020eigenstate, wang2021eigenstate, brenes2021out,  dymarsky2022bound, nussinov2022exact, jafferis2022matrix, jafferis2022jt, wang2023emergence, fava2023designs, pappalardi2023general, pappalardi2024microcanonical, fritzsch2024microcanonical}.
A recent development in this area has been the
formalization of full ETH using Free Probability Theory \cite{pappalardi2022eigenstate}. Free probability is an extension of traditional probability theory to non-commuting variables \cite{voiculescu1991limit, voiculescu1992free, speicher1997free}, which introduces powerful tools for understanding the statistical properties of large matrices, such as those that arise in chaotic quantum systems. Originating in planar field theory \cite{brezin1978planar, cvitanovic1981planar, cvitanovic1982planar}, free probability has also played a role in describing correlations in other branches of many-body physics, such as quantum information theory \cite{collins2016random}, tensor networks \cite{collins2010random, kutlerflam2021distinguishing, kudler2022negativity, cheng2024random}, disordered systems \cite{movassagh2010isotropic, movassagh2011density, chen2012error, hruza2023coherent, bauer2023bernoulli, bernard2023exact}, and gravity \cite{berkooz2019towards, penington2022replica, wang2023beyond,  wu2024non, chandrasekaran2023large}. \\
The connection between ETH and free probability provides a mathematical structure to describe correlations through noncrossing partitions, and the so-called free cumulants, connected correlation functions that generalize classical cumulants. In particular, free cumulants between observables at $2n$ times, i.e. $\kappa_{2n}(t, 0, \dots,t,0)$, provide a means to systematically study the $2n$-OTOC in Eq.~\eqref{2notoc}. 
This approach leads to the proposal that chaotic dynamics shall result at long times in ``free independence'' or \emph{freeness of observables}, which suggests that observables become statistically independent in a specific, non-classical sense as the system evolves over time \cite{cipolloni2022thermalisation, fava2023designs}.
While these ideas are mathematically appealing, their application has been primarily explored in systems evolving with random Wigner matrices \cite{cipolloni2022thermalisation} or in fine-tuned models, such as dual-unitary circuits  \cite{chen2024free}. The physical implications of long-time freeness in realistic chaotic systems remain largely unexplored.\\

In this work, we address the emergence of freeness at long times by investigating the kicked top, a paradigmatic semiclassical model of quantum chaos \cite{haake1987classical, kus1987symmetry}. 
The kicked top describes a system of spins driven periodically in time via collective interactions. 
Thanks to the latter, the system has a well-defined classical limit, corresponding to the thermodynamic limit, and the dimension of the Hilbert space scales only linearly with the system size, which makes it amenable to exact numerics.
Furthermore, this model exhibits a transition from integrability to chaos, making it an ideal setting for investigating the interplay between chaotic dynamics, freeness and its breakdown. Despite its simplicity, the kicked top captures many essential features of chaotic quantum systems, allowing us to identify several features that apply to many-body non-integrable systems. 
\\ 

Let us summarize here our main findings: 
\begin{itemize}

\item {\bf Emergence of freeness in the chaotic regime} By investigating numerically the free cumulants, \emph{we demonstrate the emergence of asymptotic freeness at long-times in the chaotic regime} of the kicked top. In contrast to integrable and mixed phase-space regions, 
where free cumulants saturate to a non-zero value, in the chaotic regime they reach a value that vanishes with the inverse of the size of the Hilbert space $D$, i.e.
$$ \kappa_{2n}(t, 0, \dots, t, 0) \sim \mathcal O(D^{-1})\, , \qquad t\gg 1 \ .$$
This provides direct evidence for the asymptotic long-time freeness of observables. 
In this regime, \emph{the long-time dynamics of the $2n$-point out-of-time-order correlators are governed by free cumulants}. 
In particular, we find at long-times
\begin{equation}
    \langle (\hat A_0(t) \hat A_0)^n \rangle \simeq \kappa_{2n}(t, 0, \dots, t, 0) \ ,
    \notag
\end{equation}
where $\hat A_0$ is a traceless observable. 
Furthermore, at long times, \emph{free cumulants decay exponentially fast} for this class of models.\\

In the chaotic regime, we further explicitly show that the $2n$-OTOC are encoded by ETH-free cumulants (defined below), which account for the matrix elements correlations. 
We also show that the ETH-free cumulants describe the leading dynamics of the square-commutator, but in this semiclassical model, they fail to reproduce its earliest time behaviour, which encodes the Lyapunov exponent.

\item {\bf Large Deviation Theory of Freeness} For systems with exponentially decaying free cumulants, we develop a \emph{large deviation approach to freeness}, inspired by standard approaches to multifractality in turbulence \cite{benzi1984multifractal,paladin1987anomalous}, chaos theory \cite{paladin1986intermittency, crisanti1988lyapunov}, and wavefunction localizations \cite{evers200fluctuations, mace2019multifractal,sierant2022universal}. When the $2n$-OTOCs decay exponentially, one can define a ``freeness time-scale'' through the quantities $\tau_n$, obtained as
\begin{equation}
    \langle (\hat A_0(t) \hat A_0)^n\rangle \sim \exp \left (- \frac{n}{\tau_n}t \right)\, \,.
    \notag
\end{equation}
$\tau_n$ are indicators to quantify how fast freeness at order $n$ is reached. 
Under a large deviation assumption, we show that these quantities shall increase monotonically with $n$ and that they shall be larger than a typical value, i.e. $$\tau_n \geq \tau_{\rm typ}\, . $$ If the times are all equal, it would indicate the absence of fluctuations and a sort of \emph{monofractal behaviour}. On the other hand, if the freeness time-scale changes with $n$, this implies a \emph{multifractal structure} of the approach to freeness. \\
We apply this large deviation perspective to the kicked top, where our numerical results reveal a hierarchy of freeness time-scale $\tau_n$ that increases with $n$, indicating a multifractal behaviour of the approach to freeness in this generic low-dimensional model of quantum chaos. We also observe the presence of a shift in time of the $2n$-OTOC only for even $n$, related to the existence of a ``Lyapunov regime'' for this class of models.

\end{itemize}

\bigskip

The rest of the paper is structured as follows. In Section~\ref{sec_mod} we recall the model of the kicked top and we introduce the main concepts of Free Probability - free cumulants and freeness - connected to the full Eigenstate Thermalization Hypothesis.
In Section~\ref{sec_dyn}, we focus on the dynamics of free cumulants; we observe the emergence of long-time freeness only in the chaotic regime
and we highlight the relevance of free cumulants 
for describing specific correlators like the $2n$-OTOCs and the square-commutator. 
In Section~\ref{sec_res}, we present the definition of a time scale for the onset of freeness, developing a large deviation theory for freeness in Section~\ref{sec_LD}. Finally, in Section~\ref{sec_num}, we study numerically the freeness time-scale, showing freeness multifractality in the model analyzed. At last, Section~\ref{sec_disc} is devoted to the conclusions and the open questions. \\ All numerical codes and datasets used in this work are openly available on Zenodo \cite{zenodo2024}.

\section{Modeling Chaos: Kicked Top, full ETH and Free Probability}
\label{sec_mod}

In this section, we recall the definition of the quantum kicked top and we provide a pedagogical review of
the Free Probability approach to the full Eigenstate Thermalization Hypothesis, which we will examine in the subsequent analysis. Readers who are familiar with these topics may proceed directly to our results in Sec.~\ref{sec_dyn} and Sec.~\ref{sec_res}.

\subsection{Kicked Top}

\begin{figure}[t]
\centering
\includegraphics[width=0.78\textwidth]{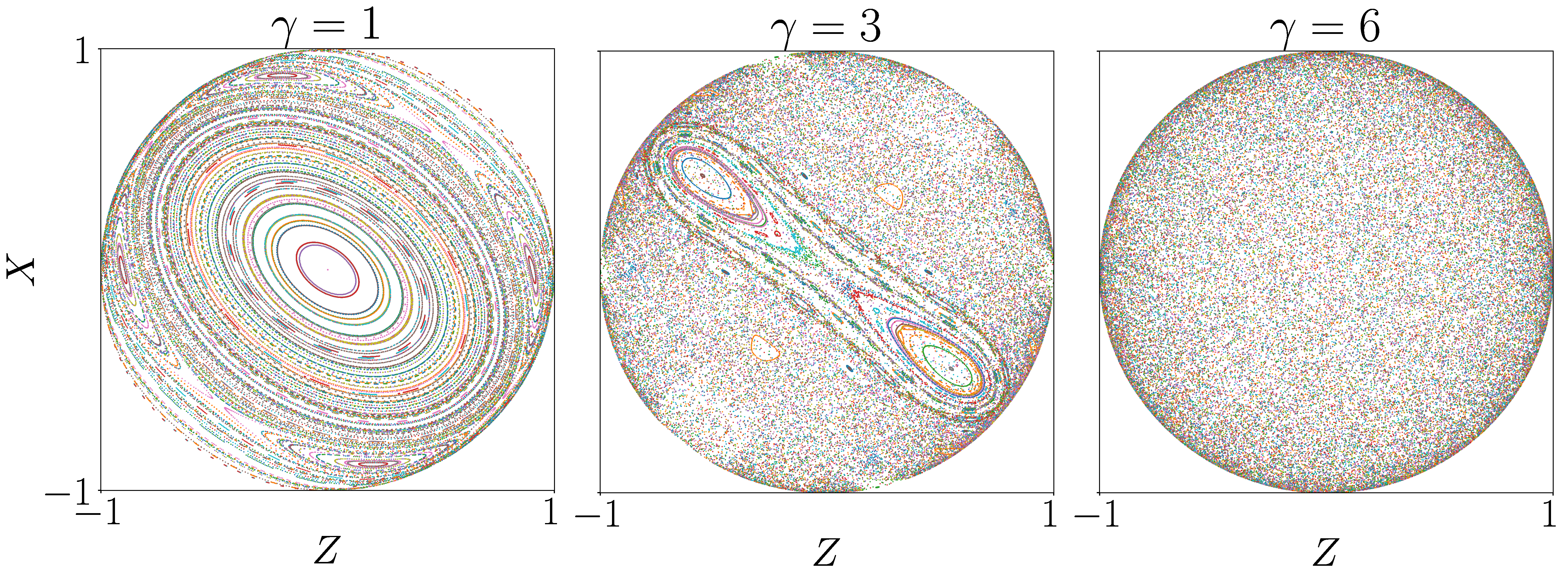}
\caption[]{Transition between regular and chaotic regimes for the classical kicked top. The classical variable is the rescaled angular momentum $(X,Y,Z)\equiv 2{(J_x,J_y,J_z)}/{N}$ which lies on a unitary sphere. Here 253 random initial conditions with $Y_0>0$ are evolved for 250 kicks through the classical equations of motion\footref{EqClassical}.}
\label{classicalk}
\end{figure}

As a paradigmatic example of semiclassical chaos, we consider a driven system: the kicked top. It is an ensemble of $N$ spins $1/2$, which is driven periodically in time via collective interactions as
\begin{equation}
\hat H(t)= p \hat J_y + \frac{\gamma}{N}\hat J^2_z \, \sum_{n=-\infty}^{\infty} \delta (t- \tau n) \ ,
\label{kickedtopHam}
\end{equation}
where $\hat J_i$ is the collective angular momentum operator in the direction $i=x,y,z$, i.e. $[\hat J_i,\hat J_j]=i\epsilon_{ijk} \hat J_k$. The parameter $p$ expresses the precession of the collective spin around the $y$ axis, and $\gamma$ represents the kicking strength. It is convenient to restrict the discussion to the total symmetric subspace, preserved by the dynamics. The latter is characterized by maximum total angular momentum $j=N/2$, fixing the Hilbert space dimension to $D=N+1$. Here $\tau$ is the period of the periodic kick, which is set to $\tau=1$.

The time evolution operator over one period is the Floquet operator
\be
 \hat U_F= e^{-\frac{i \gamma}{N}\hat J^2_z } e^{- i p \hat J_y} =  \sum_\alpha e^{-i\nu_\alpha}\ket{\nu_\alpha}\bra{\nu_\alpha}\ ,
\label{FloquetOp}
\ee
whose spectrum is given by the Floquet quasi-energies $\{\nu_\alpha\}$, which are $2\pi$ periodic; the choice $\nu_\alpha \in [-\pi,\pi), \forall \alpha$ is made.

The quantum system is characterized by a well-defined classical limit obtained for the Planck effective constant $$\hbar_{\mathrm{eff}}= \frac{\hbar}{N/2}$$ going to zero. Throughout the work, we set $\hbar=1$ such that the thermodynamic limit $N\rightarrow \infty$ corresponds to the semiclassical one.

By increasing the kicking strength $\gamma$, the system undergoes a transition between regular and chaotic dynamics: the periodic drive introduces energy into the system, making it chaotic. See Fig.~\ref{classicalk} for an illustration of the classical phase-space for different $\gamma$. Being a paradigmatic model of quantum chaos, the classical \cite{haake1987classical, kus1987symmetry} and quantum \cite{chaudhury2009quantum, wang2021multifractality} features of this transition have been extensively studied, including its OTOC dynamics \cite{pappalardi2018scrambling,  seshadri2018tripartite, sieberer2019digital, pilatowsky2020positive, lerose2020bridging}.

The kicked top, as a many-qubit system governed only by collective interactions, is studied through the effective single large-spin description. 
The investigation of long-time freeness in the presence of local interactions is left for future work.

\footnotetext{The stroboscopic classical evolution is given by the following: \begin{align*} 
X'&=(X\cos{p}+Z\sin{p})\cos{(\gamma Z')}-Y\sin{(\gamma Z')} \notag \\
Y'&=(X\cos{p}+Z\sin{p})\sin{(\gamma Z')}+Y\cos{(\gamma Z')} \notag\\
Z'&= Z\cos{p}-X\sin{p}\ .
\notag   
\end{align*}\label{EqClassical} }

\subsection{Free Probability in chaotic dynamics}

Here, we introduce the full Eigenstate Thermalization Hypothesis and the Free Probability tools - free cumulants and freeness - which can be applied to multi-time correlation functions.
Since we study a periodically driven system, we will concentrate on the description of Floquet systems, which possess a constant density of states. Thus, we will use equilibrium infinite-temperature averages
\begin{equation}
\langle \bullet \rangle = \frac {\text{tr}(\bullet )} {D}\ ,
\notag
\end{equation}
where $D$ is the dimension of the Hilbert space, and focuses on the quasi-energy eigenbasis. For discussions involving the energy eigenstates, as well as canonical and microcanonical averages, we refer the reader to Refs.~\cite{pappalardi2022eigenstate, pappalardi2023general, pappalardi2024microcanonical, fritzsch2024microcanonical}.

\subsubsection{Full Eigenstate Thermalization Hypothesis}
\label{suc_sec_fETH}
According to the Eigenstate Thermalization Hypothesis \cite{srednicki1994chaos, srednicki1999approach, dalessio2016from}, the matrix elements $A_{\alpha \beta} = \bra{\nu_\alpha} \hat{A} \ket{\nu_\beta}$ of a physical observable $\hat{A}$ in the quasi-energy eigenbasis behave as pseudorandom matrix with smooth statistical properties. The full ETH \cite{foini2019eigenstate} extends the standard formulation to multiple-point correlations between matrix elements. Specifically, statistical averages\footnote{ 
Although matrix elements are completely determined by the eigenvectors of a fixed Hamiltonian, they behave pseudo-randomly, and the statistical average shall be intended over a ``fictitious ensemble'' that can be defined heuristically in several ways,  such as averaging over a small window of energies or adding a small disordered term to the Hamiltonian and averaging over it, see e.g. \cite{deutsch1991quantum, foini2019eigenstate}. 
   In practice, the average shall be valid ``typically'' and ETH can be understood as a self-averaging property of matrix elements.} of products with distinct indices $\alpha_1 \ne \alpha_2 \ne \dots \ne \alpha_n$ read
\begin{subequations}
\label{full_ETH}
\begin{align}
\label{eth1}
\overline{A_{\alpha_1 \alpha_2}A_{\alpha_2 \alpha_3}\dots A_{\alpha_n \alpha_1}}  = D^{1-n}\, & F_{}^{(n)} \left(\omega_{\alpha_1\alpha_2},\dots, \omega_{\alpha_{n-1}\alpha_n}\right),
\end{align}
where $F^{(n)}$ are some smooth functions of the eigenenergies differences $\omega_{\alpha_1\alpha_2}=\nu_{\alpha_1}-\nu_{\alpha_2}$ that define the observable. 
For $n=1,2$ one recovers the standard ETH \cite{srednicki1999approach}, where in the Floquet basis $F^{(1)} = \langle \hat A \rangle$ is the constant equilibrium average and $F^{(2)}(\omega)=|f(\omega)|^2$ is the dynamical correlation function. When indices are repeated, the average of the products factorizes at the leading order in $D$ as
\begin{align}
    \overline{A_{\alpha_1 \alpha_2} \dots A_{\alpha_{m-1} \alpha_1} A_{\alpha_1 \alpha_{m+1}} \dots A_{\alpha_n\alpha_1} } \label{eth2} = \overline{A_{\alpha_1 \alpha_2} \dots A_{\alpha_{m-1} \alpha_1}}\,\, \overline{ A_{\alpha_1 \alpha_{m+1}} \dots A_{\alpha_n\alpha_1} } + & \mathcal{O}\left(D^{-1}\right)\, .
\end{align}
\end{subequations}

The full ETH ansatz is needed to study equilibrium multi-time correlation functions, where products of many-matrix elements naturally appear, i.e. 
\begin{align}
\langle & \hat{A}(t_1) \dots  \hat{A}(t_n) \rangle = \frac{1}{D} \sum_{\alpha_1, \dots, \alpha_n}  e^{i \vec \omega \cdot \vec t} A_{\alpha_1 \alpha_2}\dots A_{\alpha_n \alpha_1} \label{correlation4} 
\\ & \text{with}\quad \vec \omega=(\omega_{\alpha_1 \alpha_2}, \dots,\omega_{\alpha_{n} \alpha_1}) \ ,\quad \vec t=(t_1,\dots,t_n) \notag
\end{align}
As a result of the smoothness of the matrix elements and of the scaling with $D$ in Eq.~\eqref{full_ETH},  Ref.~\cite{foini2019eigenstate} showed that 
the correlation functions at each $n$ are organized in terms of sums over distinct indices as
\begin{equation}
    \mathcal{K}_n(\vec t \,) := \frac{1}{D} \sum_{\alpha_1\ne\dots \ne \alpha_n}  e^{i \vec \omega \cdot \vec t}  A_{\alpha_1 \alpha_2} \dots A_{\alpha_n \alpha_1} \ .
    \label{freeceth}
    \end{equation} 
These are referred to as \emph{ETH simple loops} because, in the diagrammatic representations of the full ETH, they correspond to diagrams where energy indices appear as dots arranged in a single loop. For completeness, a detailed discussion of ETH diagrams and the derivation of this result is provided in App.~\ref{app_fETH}.
Furthermore, according to Eq.~\eqref{eth1}, the simple loops can be written as 
    $$\mathcal{K}_n(\vec t\,)= \int d \vec{\omega}  e^{i \vec \omega \cdot \vec t} F^{(n)}(\vec \omega)\ ,$$ which states that ETH smooth functions are linked to the Fourier Transform of $\mathcal{K}_n(\vec t\,)$.
In summary, \textit{within ETH the building blocks of multi-time correlations are given by simple loops $\mathcal{K}_n(\vec t \,)$.}

\subsubsection{Free Cumulants}

The simple loops $\mathcal{K}_n$ in Eq.~\eqref{freeceth} appearing in ETH description of multi-time correlators can actually be identified as \emph{free cumulants}, a familiar object in the theory of Free Probability \cite{speicher1997free}, see  Refs.~\cite{speicher2016free, xia2019simple} for an introduction.

Free cumulants $\kappa_n$ are connected correlation functions defined from moments of $n$ variables. Let us here consider the case of a time-dependent observable at different times: $\hat A(t_i)$ $i=1, \dots, n$. Free cumulants are defined implicitly and recursively through the moments-free cumulant formula \cite{speicher1997free}:

\begin{align}
\label{def_Free}
\langle \hat A(t_1)\dots \hat A(t_n)\rangle & = \sum_{\pi \in NC(n)} \kappa_\pi(\hat A(t_1), \dots, \hat A(t_n))\ ,
\end{align}

where $\pi \in NC(n)$ are noncrossing partitions of a set of $n$ elements. The $\kappa_\pi$ is the product of free cumulants, one for each block $B$ of the partition, based on the number of elements of the block, i.e.
\begin{equation}
    \kappa_\pi(\hat A(t_1), \dots, \hat A(t_n)) = \prod_{B\in \pi} \kappa_{|B|} \left(\prod_{q=1}^{|B|} \hat{A}(t_q)\right)\ ,
\end{equation}
where
$q$ counts the number of terms in each block from 1 to its length $|B|$. 
As an example,  the first free cumulants for equal-times observables $\hat A(t_i)= \hat A$ are defined by
\begin{align}
\langle \hat A \rangle&=\kappa_1 \notag\\
\langle \hat A^2 \rangle&=\kappa_2+\kappa^2_1 \notag\\
\langle \hat A^3 \rangle&= \kappa_3+ \kappa^3_1+ 3\kappa_1\kappa_2 \label{examplescum}\\
\langle \hat A^4 \rangle&= \kappa_4+ \kappa^4_1 + 6 \kappa^2_1\kappa_2 + 4 \kappa_1 \kappa_3+2\kappa^2_2 \notag
\end{align}
with the notation $\kappa_n\equiv \kappa_n(\hat A, \dots, \hat A)$. 
\footnote{The definition is based on the combinatorial aspects of partitions of a set of $n$ elements; in the same way as the classical cumulants are related to the lattice of all partitions, free cumulants are related only to the lattice of noncrossing partitions. 
In the example \eqref{examplescum}, cumulants for the classical theory of probability would have been defined in the same way apart from $\langle A^4 \rangle_{cl} = \langle A^4 \rangle +\kappa^2_2$, for the contribution of one crossing partition (there are no crossing partitions for $n<4$).}

While Eq.~\eqref{def_Free} is just an implicit definition of free cumulants in terms of moments, when ETH applies, the \emph{free cumulants correspond to the ETH simple loops $\mathcal{K}_n$} defined in Eq.~\eqref{freeceth} namely \cite{pappalardi2022eigenstate} 
\begin{equation}
\kappa_n \left(\hat{A}(t_1),\dots, \hat{A}(t_n)\right) = \mathcal{K}_n (\vec{t}\, ) + \mathcal O(D^{-1}) \ .
\label{identcum}
\end{equation}

This follows from the ETH properties (1,2) discussed in App.~\ref{app_fETH}, which show that multi-time correlations have exactly the combinatorics of noncrossing partitions and that they can be rewritten only in terms of simple loops.
In the example of $n=4$, the free cumulants are defined through Eq.~\eqref{examplescum}, while the expression of the equal-time correlation function through ETH is given in Fig.~\ref{pictorial4corr} in the App.~\ref{app_fETH}; these two formulas have the same structure. This can be generalized $\forall n$, allowing the identification of free cumulants $\kappa_n$ with ETH simple loops $\mathcal{K}_n$, from now on called \textit{ETH-free cumulants}. Therefore, free probability provides the combinatorics needed for the calculation of multi-time correlations, when ETH is applied. 

\subsubsection{Freeness}

A central concept in Free Probability is free independence, or \emph{freeness}, which generalizes the classical notion of independence from commuting to non-commuting variables. Two large $D \times D$ matrices $A$ and $B$ are said to be \emph{asymptotically free} if all their mixed free cumulants (defined in Eq.~\eqref{def_Free}) vanish $\forall n$:

\begin{equation}
    \label{free_def}
   \kappa_{n}(A,B,\dots )=0 \ ,
\end{equation} 
where ``asymptotically'' indicates that this condition holds as the matrix dimension $D$ diverges. The dots denote additional arguments, the term \textit{mixed} indicates that at least one entry differs from the others, without any constraint on their arrangement. When $A$ and $B$
are free, their mixed moments can be computed from their individual moments only\,\footnote{For example, consider the implicit definition in Eq.~\eqref{def_Free} for alternating free variables $A, B$:
\begin{align*}
    \langle AB\rangle & = \kappa_1(A)\kappa_1(B) + \kappa_2(A, B)   \qquad  \text{with}\qquad \kappa_2(A, B) = 0
    \\  \to \langle AB\rangle & = \langle A\rangle \langle B \rangle\ , 
    \\
    \\
    \langle ABAB\rangle  & =  \kappa_4  (A, B, A, B) + \kappa^2_1(A)\kappa^2_1(B) 
    + 4 \kappa_1(A)\kappa_1(B)\kappa_2(A, B) 
    \\ & + \kappa^2_1(A) \kappa_2(B) + \kappa^2_1(B) \kappa_2(A)
      + 2 \kappa_1(A) \kappa_3(B,A,B) + 2 \kappa_1(B) \kappa_3(A,B, A)
      \\ & + 2 \kappa^2_2(A,B)\quad \text{with}\quad \kappa_n(A, B, \dots) = 0
    \\ \to  \langle ABAB\rangle  & =  \left(\langle A^2\rangle -\braket A^2 \right) \braket{B}^2 + \braket{A}^2 \braket{B^2}\ .
\end{align*}
}. 

If the vanishing of free cumulants in Eq.~\eqref{free_def} holds only up to a certain order $n$, then $A$ and $B$ are termed \emph{asymptotically $n$-free}. Important examples of free variables can be found in random matrix theory \cite{voiculescu1991limit, mingo2017free}, specifically when considering randomly rotated deterministic matrices $B$ and $A^U = U^{\dagger}AU$. When $U$ is sampled from the Haar ensemble, $A^U$ and $B$ are asymptotically free \cite{mingo2017free}. 
If $U$ is sampled from an $n$-design (an ensemble of unitaries reproducing moments of the Haar measure up to order $n$), then $A^U$ and $B$ are asymptotically $n$-free \cite{fava2021hydrodynamic}, though the converse is not generally true; the precise relation between approximate designs and approximate freeness remains an active area of research~\cite{dowling2025free}.  \\

The definition of freeness naturally leads to the question of whether chaotic time evolution results in asymptotic $n$-freeness of physical observables at long times, in the thermodynamic limit. Such a behaviour would suggest that chaotic time evolution at infinite times is as effective as a $n$-design in inducing freeness among observables.
While this holds for random matrices \cite{cipolloni2022thermalisation}, it shall follow  
for generic many-body Hamiltonians as a consequence on the Eigenstate Thermalization Hypothesis \cite{fava2023designs}. Indeed, for quantum observables $\hat{A}(t)$ and $\hat{B}$ satisfying ETH, the infinite-time average of their mixed free cumulants vanishes with the Hilbert-space dimension, when taking into account finite-size effects:
\begin{equation}
    \label{eq_infinity}
    {[\kappa_{n}]}_\infty \equiv \lim_{T\to \infty} \frac 1T \int_0^T dt \,\,\kappa_{n}(\hat{A}(t),\hat{B}, \dots ) = {\mathcal O(D^{-1}})\ .
\end{equation}

Furthermore, the infinite-time average of the temporal fluctuations are also suppressed:
\begin{equation}
\label{eq_infinity_2}
    {[\kappa_{n}^2]}_\infty -{[\kappa_{n}]}^2_\infty \sim \mathcal O(D^{-2})\ .
\end{equation}
These results are derived using i) the ETH, which expresses free cumulants as sums of simple loops, and ii) a no-resonance condition between eigenvalues, believed to hold generically in non-integrable systems \cite{srednicki1999approach, linden2009quantum}, see App.~\ref{appB} for details on the proof.

Since the temporal fluctuations of the expectation value of the observable are suppressed with system size, in the large $D$ limit, at almost all times, the free cumulants $\kappa_{n}(t)$ vanish.\\
In a physical system, it is conjectured that asymptotic $n$-freeness is reached at a finite long-time, that is 
\be
\kappa_{n}\left(\hat{A}(t),\hat{B},\dots, \right) \approx \mathcal O(D^{-1}) \, , \quad \quad \text{for}\quad  t \gg 1\ .
\label{freenessthermal}
\ee
This should be understood as an approximate notion of freeness, defined in the long-time limit and asymptotically in the system size. \\
Eqs.\eqref{eq_infinity}-\eqref{freenessthermal} are a consequence of ETH, hence they remain, up to now, an hypothesis. 
One of the goals of this work is to investigate numerically whether this regime is reached, and to properly define the time scale associated with freeness, which generically depends on $n$ and that shall quantify how fast freeness is achieved.\\
In general, the calculation of the free cumulants follows from the recursive definition in Eq.~\eqref{def_Free}. 
To simplify the calculations, we will often consider traceless observables, i.e. 
\begin{equation}
 \hat{A}_0=\hat{A}- \langle \hat{A} \rangle \mathbf{I}\, ,
\label{eq:A0}
\end{equation}
in order to have $\kappa_1(\hat A_0) = \langle \hat{A_0} \rangle=0$. The free cumulants of $\hat A_0$ can be recursively computed starting from 
$\kappa_1 (\hat A_0)= \langle \hat{A}_0 \rangle=0$, as 
\begin{subequations}
\label{calculate_free}
\begin{align}
    \kappa_2(t,0) & =\langle \hat{A}_0(t)\hat{A}_0  \rangle\ , 
    \\ 
    \kappa_3(t,0,t)& =\langle \hat{A}_0(t)\hat{A}_0\hat{A}_0(t)  \rangle\ , 
    \\ 
    \kappa_4(t,0,t,0)& = \langle \hat{A}_0(t)\hat{A}_0\hat{A}_0(t) \hat{A}_0 \rangle -2\kappa^2_2(t,0)\ ,\label{k4}
    \\ 
    \kappa_5(t,0,t,0,t)& = \langle \hat{A}_0(t)\hat{A}_0\hat{A}_0(t) \hat{A}_0 \hat{A}_0(t) \rangle -5\kappa_2(t,0) \kappa_3(t,0,t)\ ,
    \\ 
    \kappa_6(t,0,t,0,t,0)& =\langle \hat{A}_0(t)\hat{A}_0\hat{A}_0(t) \hat{A}_0 \hat{A}_0(t) A_0\rangle \notag  -5\kappa^3_2(t,0)+3\kappa^2_3(t,0,t)  \notag  \\ & \quad-6\kappa_2(t,0)\kappa_4(t,0,t,0)\ ,
\end{align}
\end{subequations}
and so on. The higher-order cumulants are evaluated via symbolic computation; the expressions can be found in \cite{zenodo2024}.

\paragraph{Alternating free cumulants}
Among all possible mixed free cumulants, we focus on alternating free cumulants, namely those with alternating arguments in Eq.~\eqref{free_def}, $\kappa_{2n}(A,B,\dots,A,B)$. When $A$ and $B$ are free, the alternating moments admit a compact expression, given by \cite{mingo2017free} 
\begin{equation}
\label{eq_free_prod}
\langle (A B )^n\rangle
=\sum_{\pi \in NC(n)} \kappa_\pi(A, \dots, A) \, \langle B^n \rangle_{\pi^*} \ ,
\end{equation}
where $\langle  B^n\rangle_{\pi^*}$ is the product of moments, one for each term of the partition $\pi^*$, known as the Kraweras complement (or dual) of the partition $\pi$ \footnote{The Kraweras complement $\pi^*$ of a partition $\pi$ is defined by the property that the blocks composing $\pi^*$ are the maximal blocks (polygons) with vertices on  ``$B$'' that do not cross the blocks of $\pi$.}.\\
In this work we want to probe asymptotic freeness through the study of \emph{free cumulants of out-of-time order correlators} such as 
\begin{equation}
    \kappa_{2n}(t, 0, \dots, t, 0) \equiv \kappa_{2n}(\hat A(t), \hat A, \dots, \hat A(t), \hat A)\ \notag .
\end{equation} 
This choice is motivated by the fact that they are related to $2n$-OTOC, which are of physical interest, and by the fact that the alternating free cumulants exhibit the slowest dynamics between all the mixed free cumulants and, consequently, determine the freeness scale at order \(n\). Numerical justifications of this statement can be found in App.~\ref{app_C}.\\ 
Long-time asymptotic freeness implies that out-of-time order correlators at late times can be computed using established properties of Free Probability \cite{mingo2017free} such as the factorization of Eq.~\eqref{eq_free_prod}.
For instance, the OTOC between $\hat{A}(t)$ and $\hat{B}$ at long-times $t \gg 1$ is given by:
\begin{equation}
    \langle \hat{A}(t) \hat B \hat{A}(t) \hat B\rangle 
    = \kappa_2(\hat{A},\hat{A}) \langle \hat B\rangle ^2 + \langle \hat{A} \rangle^2 \langle \hat B^2\rangle + \mathcal O(D^{-1})\ .
    \label{fac}
\end{equation}
In the case of traceless observables $\langle \hat A\rangle = \braket{\hat B}=0$, this quantity vanishes as $\mathcal O(D^{-1})$.

\section{Free cumulants dynamics}
\label{sec_dyn}

Let us now present our results for the dynamics of alternating free cumulants in the kicked-top model.\\
To study the dynamics at stroboscopic times, we perform an exact diagonalization of the Floquet operator given in Eq.~\eqref{FloquetOp}, with the parameter fixed at $\displaystyle p = 4\pi/7$. Since the Hamiltonian in Eq.~\eqref{kickedtopHam} is invariant under $\pi$-rotations around the $y$-axis, we restrict our analysis to the positive-parity subspace, which has a dimension of $D = N/2 + 1$ when $j=N/2$ is even\footnote{When $j=N/2$ is odd the dimension is $D = N/2$, while if $j=N/2$ is half-integer the dimension is $D=\frac{1}{2}(N+1)$.  }. For concreteness, we focus on the dynamics of the observable
\begin{equation}
    \hat{A} = \frac{\hat{J}_y^2}{(N/2)^2} \, ,
\label{eq:obs}
\end{equation}
which has a finite average of $\displaystyle \kappa_1 = \langle \hat{A} \rangle = 1/3 + \mathcal{O}\left(D^{-1}\right)$.  We have confirmed that the qualitative behaviour of our results is robust to this choice of observable. Indeed, being the kicked top a system in which locality plays no role, all the collective spin observables, apart from the conserved quantity, display the same dynamics for the $2n-$OTOCs and the free cumulants, up to observable-dependent prefactors, and satisfy ETH in the chaotic regime.\\

We begin by discussing long-time asymptotic freeness in the regular, mixed and chaotic regimes (Sec.~\ref{sec_tran}). Secondly, we focus on the chaotic limit, and we show that the dynamics at long times of $2n$-OTOCs of traceless observables is given only by the highest-order free cumulant $\kappa_{2n}(t,0,\dots,t,0)$ (Sec.~\ref{sec_OTOCfree}). Finally, we confirm the validity of ETH in the chaotic regime, showing explicitly the decomposition of multi-time correlation functions in  ETH-free cumulants; moreover, we pinpoint the inadequacy of the ETH prediction when it comes to describing the early-time behaviour of the square-commutator (Sec.~\ref{sec_ETH}). 

\subsection{Long-time freeness and chaos}
\label{sec_tran}

\begin{figure}[t]
\centering
\includegraphics[width=1\textwidth]{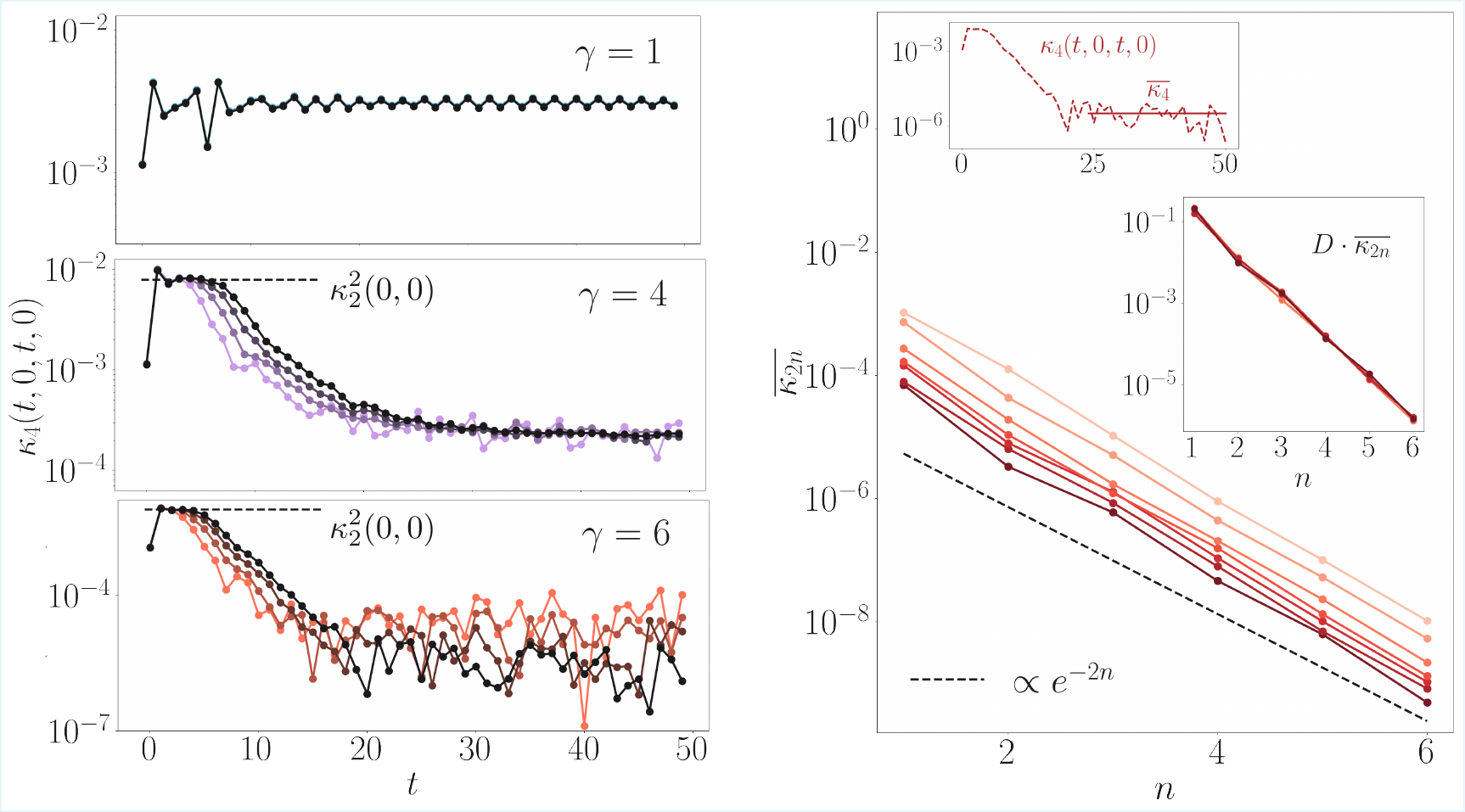}
\caption{Numerical study for investigating the presence of asymptotic long-time freeness in the kicked top \eqref{kickedtopHam}, for the traceless observable $\hat A_0$ obtained from $\hat A$ in Eq.~\eqref{eq:obs}.\\
(left) Evolution in time of $\kappa_4(t,0,t,0)$. Three regimes are displayed: regular for $\gamma=1$ in the upper panel, mixed for $\gamma=4$ in the central panel and fully chaotic for $\gamma=6$ in the bottom panel. 
In every panel, different system sizes are shown ($N \in [600,1400,3000,6000]$,  from light to dark colors).  \\
(right) Saturation value at long-times of the free cumulants in the chaotic case $\gamma=6$, as a function of $n$, for different system sizes ($N \in [300,600,1400,2000,3000,4000,6000]$,  from light to dark colors).
$\overline{\kappa_{2n}}$ is obtained averaging $\kappa_{2n}(t,0,\dots,t,0)$ between $t=25$ and $t=50$ as shown in the inset on the left, for $n=2$.}
\label{fig_1}
\end{figure}

We find that long-time asymptotic freeness is achieved only in the regime associated with full chaos in the classical limit. This is evident already at the level of the fourth-order free cumulant $\kappa_4(t, 0, t, 0)$, expected to vanish in the presence of asymptotic freeness for large $D$, as ${\kappa_4(t, 0, t, 0) \sim \mathcal{O}(D^{-1})}$. In Fig.~\ref{fig_1}(left), we report the numerical evaluation of the free cumulant, calculated using Eq.~\eqref{k4}, for different system sizes across three regimes: regular (upper panel, $\gamma=1$), mixed (center panel, $\gamma=4$), and fully chaotic (bottom panel, $\gamma=6$). The data show that in the regular and mixed phase-space regimes, the saturation value remains constant with increasing system size, indicating the absence of long-time freeness. In contrast, for $\gamma=6$, at long times the free cumulant fluctuates around a value that decreases as the Hilbert space dimensionality $D$ increases, as predicted by Eq.~\eqref{freenessthermal} for $\hat{B}=\hat{A}(0)$. The absence of late-time freeness in the regular and mixed phase-space can be attributed to the failure of the Eigenstate Thermalization Hypothesis, which is instead satisfied in the chaotic regime, as we discuss in more detail in Sec.~\ref{sec_ETH}.\\
As we discuss in Sec.~\ref{sec_num}, we find that the decay towards the plateau in the chaotic and mixed phase occurs after the Ehrenfest time.
It is interesting to see what happens before this time, i.e. in the classical regime. We observe that at short times, the free cumulant reaches a constant value given by $\kappa_2(0,0)\kappa_2(t,t) = \kappa^2_2(0,0)$. This value is indicative of classical independence\footnote{Independence between two commuting (classical) variables $x, y$ is referred to as ``classical independence'' and it is associated by the vanishing of all joint (classical) mixed cumulants \cite{speicher2025lecturey}, i.e. $\kappa_n^{\rm classical}(x, y, \dots)=0$.}: indeed, the fourth free cumulant differs from the classical fourth cumulant by precisely this factor, coming from the crossing partition $\kappa_4(t,0,t,0) =\kappa^{\rm classical}_4(t,0,t,0) + \kappa_2^2(0,0) $. Under classical independence the fourth classical cumulant vanishes, so that $\kappa_4 \approx \kappa_2^2$.
This happens till the Ehrenfest time, when the finite-dimensional semiclassical system exhibits classical behaviour: in the first few steps the quantum states spreads all over the Hilbert space,  compatibly with classical chaos, and shows classical independence. However, as time progresses, quantum effects emerge, leading to a gradual departure from classical independence, and eventually to the onset of freeness.
These two conditions, classical independence and freeness, indeed, cannot coexist, but instead appear sequentially in time. A similar behaviour is observed also in the mixed regime, where classical independence is present even in the absence of freeness. It would be worthwhile to investigate this aspect further, and to better understand the value of the final plateau. We leave this task for future work.

To quantitatively characterize the emergence of asymptotic freeness in the chaotic regime,
we focus on $\gamma=6$, and we analyze the long-time saturation value of the free cumulants $\kappa_{2n}(t,0\dots,t,0)$ as $n$ and the system size $D$ are changed. We compute the exact time evolution, using Eq.~\eqref{calculate_free}, up to $n=6$ \footnote{Numerically, for evaluating the free cumulant $\kappa_{2n}$, we calculate the respective $2n-$OTOC and subtract all the products of powers of lower-order free cumulants, following exactly the moments-cumulants \eqref{eq:momcum}.
The exact expressions for all the orders up to $n=6$, for traceless observables, can be found in \cite{zenodo2024}. \label{foot_num_cum}}.
In Fig.~\ref{fig_1}(right) we consider the time-average $\overline{\kappa_{2n}}$, obtained averaging $\kappa_{2n}(t,0,\dots,t,0)$ over time, between $t=25$ and $t=50$. Free cumulants saturate to a value that vanishes as $D^{-1}$, as encoded in Eq.~\eqref{eq_infinity}. This is a known feature characterizing long-time dynamics of out-of-time ordered correlators \cite{huang2019finite}. 
In the case of free cumulants as well, this value stems from the fluctuations of matrix elements, and as such, it is not presently captured by the full ETH in Eqs.~\eqref{full_ETH}. \\ 
The data show that the plateau is also decreasing with the order of the power $n$, proportionally to $e^{-2n}$.
While the scaling with
$D$ originates from the fluctuations of the matrix elements and is universally valid for all observables obeying the ETH, the scaling with $n$ can be written as $a^n$, where the value of $a<1$ is observable-dependent and unrelated to asymptotic freeness; this feature, indeed, does not impact the general validity of the subsequent results, which are observable independent.
Overall, we find 
\begin{equation}
    \overline{\kappa_{2n}} \sim \frac{a^n}{D} \ .
    \label{eq:plateau}
\end{equation}
Since it will enter the discussion below, one should notice that, as an immediate consequence of Eq.~\eqref{eq:plateau}, products of $\alpha$ free cumulants scale as $D^{-\alpha}$.
After the saturation to $D^{-1}$, one may eventually expect signatures of level repulsion at even longer time scales, see e.g. Ref.~\cite{cotler2017chaos, cotler2020spectral}. A more detailed understanding of this phenomenon is left to future investigation.

\begin{figure}[t]
\centering
\includegraphics[width=0.7\linewidth]{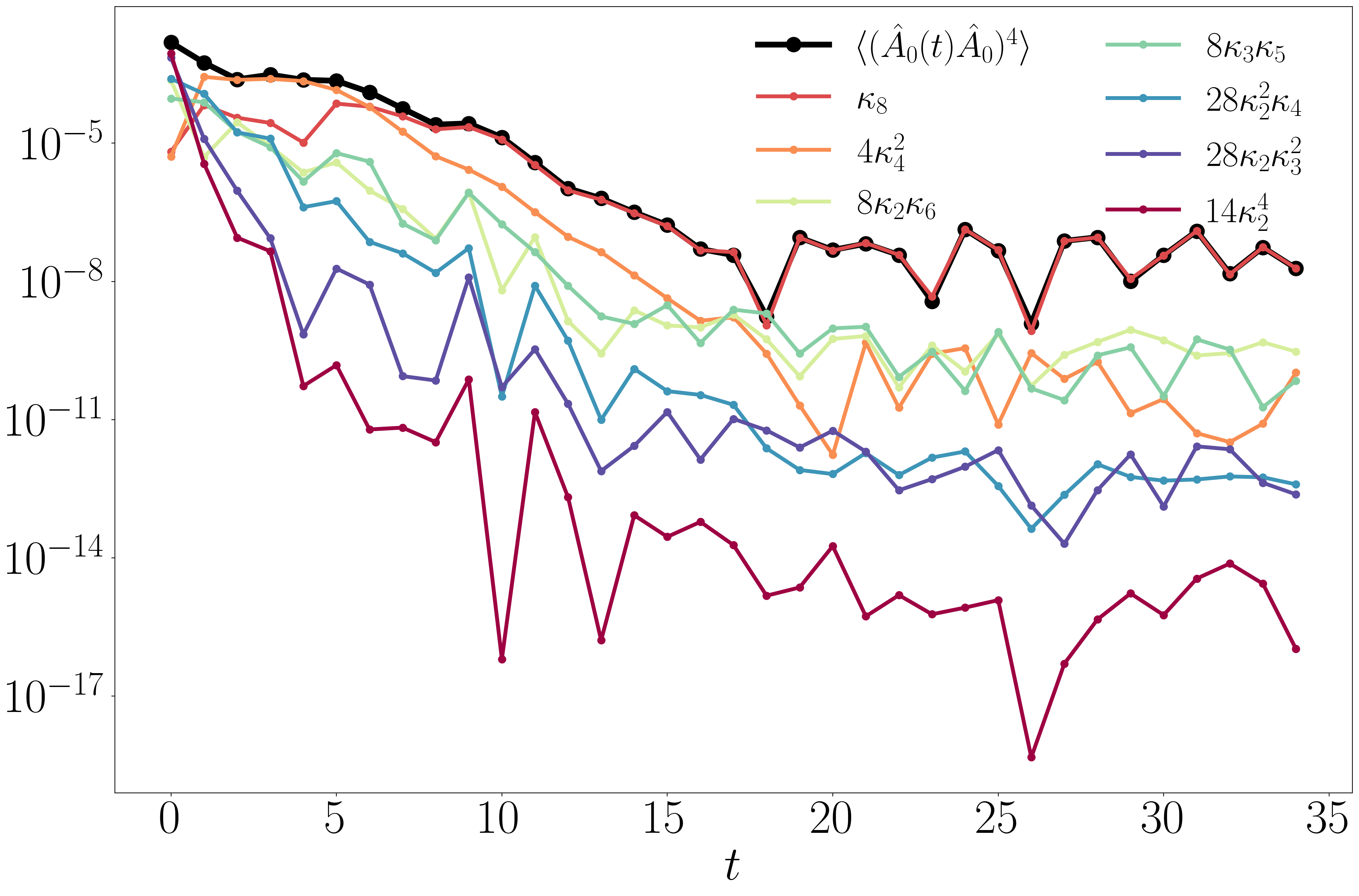}
\caption{Dynamics of the 8-OTOC compared with the free cumulants in which it can be rewritten (Eq.~\eqref{4OTOC}). The traceless observable $\hat A_0$ is obtained from $\hat A$ in Eq.~\eqref{eq:obs}, for the kicked top \eqref{kickedtopHam} in the chaotic case $\gamma=6$, for $N=6000$. The generalized Ehrenfest time for these parameters is around $t^{(4)}_{\rm Ehr} \sim 6$.
}
\label{OTOCnumerical}
\end{figure}

\subsection{Free cumulants and $2n$ out-of-time order correlators} 
\label{sec_OTOCfree}

Free cumulants play a crucial role in the dynamics of the $2n$-out-of-time order correlation $\langle (\hat{A}(t)\hat{A})^n \rangle$, as shown in Refs.~\cite{pappalardi2023general, fritzsch2021eigenstate} for the standard OTOC at $n=2$. We will now discuss it for an arbitrary order of $n$.   
 In particular, we focus on the case $n=4$, but the same results have been obtained for different values of $n$.
We consider $2n$-OTOCs of traceless observables, as in Eq.~\eqref{eq:A0}; we will restore the effect of a finite first moment in Sec.~\ref{sec_ETH} below. Having traceless operators $\langle \hat{A_0} \rangle=0$ implies that the $2n$-OTOCs should asymptotically vanish to zero over a long-time since the time-independent plateau has been shifted to zero, see the right-end side of Eq.~\eqref{fac}.

 In Fig.~\ref{OTOCnumerical}, we compare the evolution of the 8-OTOC, with the  various free cumulants of $\hat A_0$ which appear in its decomposition, namely 
 \begin{align}
\langle (\hat{A}_0(t)\hat{A}_0)^4 \rangle& = \kappa_8(t,0,\dots,t,0) + 4\kappa^2_4(t,0,t,0)+ 8\kappa_6(t,0,t,0,t,0)\kappa_2(t,0) +\notag \\ &+ \kappa_5(t,0,t,0,t)\kappa_3(t,0,t)\notag + 28 \kappa^2_2(t,0)\kappa_4(t,0,t,0)+ \\&+28 \kappa_2(t,0)\kappa^2_3(t,0,t) +14\kappa^4_2(t,0)\ .
\label{4OTOC}    
\end{align} 
We observe that at long times, the evolution of the 8-OTOC is given only by $\kappa_{8}(t,0,\dots,t,0)$, while the other contributions decay to zero more rapidly.
In other words, the higher free cumulant constitutes the biggest contribution to the $2n$-OTOC and one can identify the two as
\begin{equation}
    \langle (\hat A_0(t) \hat A_0)^n \rangle \sim \kappa_{2n}(t, 0, \dots, t, 0) \quad \text{for} \quad t\gg 1\ .
    \label{eq:momcum}
\end{equation}

At long times, where the free cumulants reach a plateau, this feature is compatible with the scaling observed in Eq.~\eqref{eq:plateau} for which products of $\alpha$ free cumulants saturates to $D^{-\alpha}$. Hence, $\kappa_{2n}$, being the only contribution saturating to $D^{-1}$, dominates over the other smaller terms and gives the plateau value of the $2n$-OTOC.  
\\Eq.~\eqref{eq:momcum} is also valid in an intermediate regime at sufficiently long-times.  Furthermore, the numerical results in Fig.~\ref{OTOCnumerical} show that for this model freeness is reached exponentially fast 
\begin{equation}
\langle (\hat A_0(t) \hat A_0)^n \rangle \sim \kappa_{2n}(t, 0, \dots, t, 0) \sim \text{exp}\left(- \frac{n}{\tau_n} t \right) \, .
\end{equation}
The quantities $\tau_n$ so defined,
\begin{equation}
    \frac 1{\tau_{n}} = - \frac{1}{nt} \log \left|\langle (\hat A_0(t) \hat A_0)^{n}\rangle \right| 
    \quad \text{for}\quad t\gg 1 \, ,
\label{defTauN} 
\end{equation}
indicate how quickly freeness is achieved and can be referred as definition of a \textit{freeness time-scale}. Indeed, it provides an estimate of the time at which $\kappa_{2n}$ reaches its $n$-dependent plateau $a^n$.
 We note that standard $4$-OTOC was already found to decay exponentially fast after the Ehrenfest time in chaotic single-body quantum systems with classical limits, see e.g. Refs.\cite{garcia2018chaos, notenson2023classical}. As we discuss in Sec.~\ref{sec_num}, we find that the exponential decay occurs after the  (generalized) Ehrenfest time for all even free cumulants.
The goal of Sec.~\ref{sec_res} will be to interpret the decay rate for arbitrary $n$ in a general way, from a large deviation theory.

\subsection{ETH-free cumulants in the chaotic regime}
\label{sec_ETH}

At the beginning of this section, we stated that asymptotic freeness is achieved in the chaotic regime as a result of the validity of the Eigenstate Thermalization Hypothesis. Focusing on the regime for $\gamma=6$, we now demonstrate how the correlations from ETH serve as the fundamental building blocks for multi-time correlators in the large $N$ limit.
For this purpose, we will consider the evolution of the observable $\hat A$ in Eq.~\eqref{eq:obs} which possesses finite average $\displaystyle \kappa_1 = \langle \hat{A} \rangle \simeq 1/3$.

\begin{figure}[t]
\centering
\includegraphics[width=0.7\linewidth]{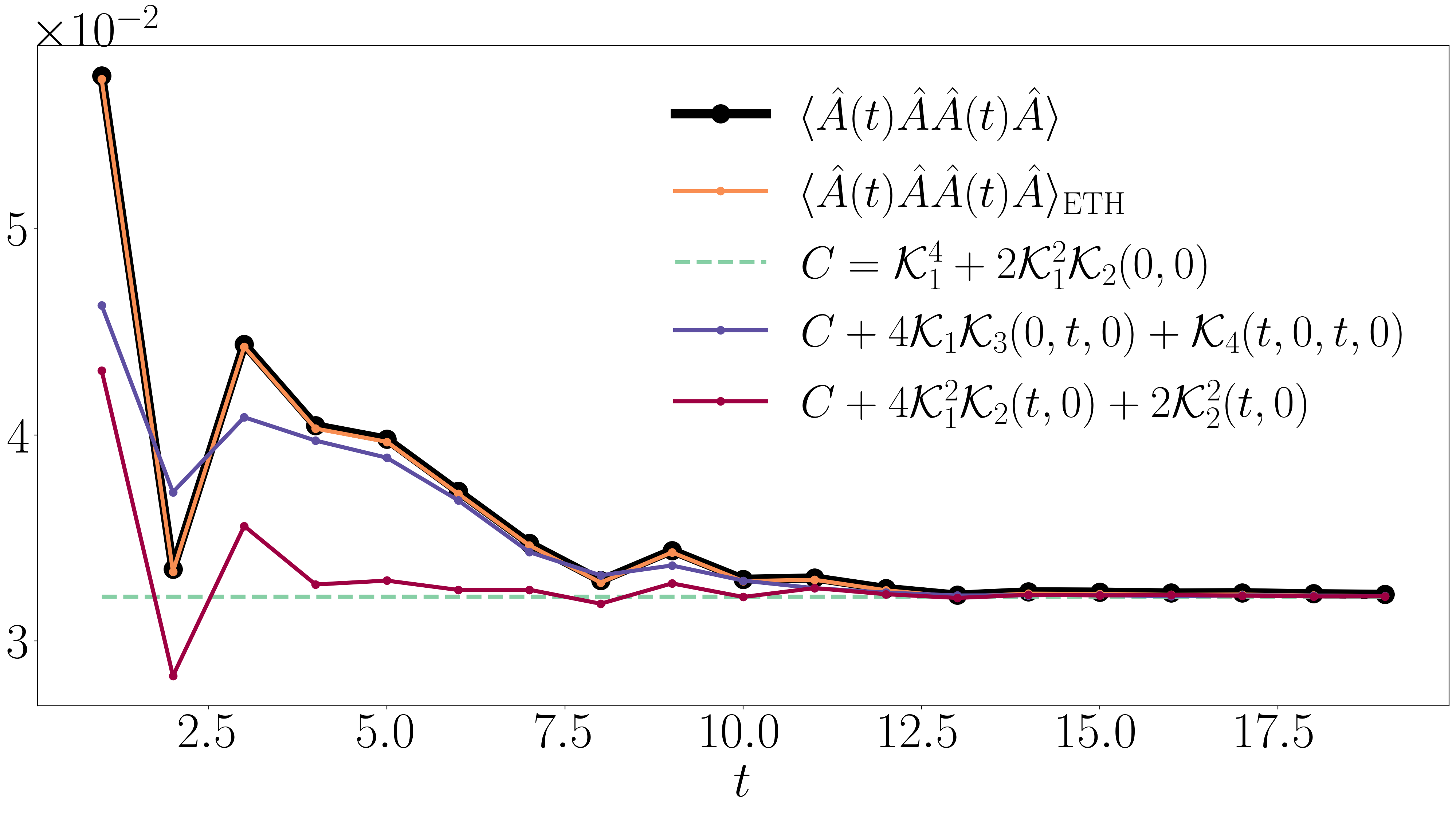}\caption{Dynamics of the OTOC in Eq.~\eqref{OTOCexact} compared with the one calculated through the ETH decomposition in Eq.~\eqref{OTOCfreeprob}. The contributions from different free cumulants are displayed with different colours. The observable chosen is $\hat A$ in Eq.~\eqref{eq:obs}, for the kicked top \eqref{kickedtopHam} in the chaotic case $\gamma=6$, for $N=6000$. } 
\label{fullETHOTOC}
\end{figure}

\paragraph*{OTOC decomposition and ETH-free cumulants}
We start by showing that the building blocks of the dynamics are given by the ETH simple loops, in other words, that the free cumulants
correspond at the leading order to sums over different indices, as stated in Eq.~\eqref{identcum}, that is
\begin{equation}
    \kappa_n(t, 0, \dots ,t,0 ) = \mathcal K_n(t, 0, \dots,t,0, ) + \mathcal O(D^{-1}) \, .
    \notag
\end{equation}
We study the exact dynamics of the out-of-time order correlator
\begin{align}
\langle \hat{A}(t)\hat{A}\hat{A}(t)\hat{A}\rangle = \frac{1}{D} \sum_{i,j,k,l} e^{i (\omega_{ij} + \omega_{kl} )t} A_{ij} A_{jk} A_{kl} A_{li}\ ,
\label{OTOCexact}
\end{align}
and contrast it with the prediction obtained via ETH, namely:
\begin{align}
\langle \hat{A}(t)\hat{A}\hat{A}(t)\hat{A} \rangle_{\mathrm{ETH}} & = \mathcal{K}_4(t,0,t,0) + \mathcal{K}_1^4  +  4\mathcal{K}_1^2 \mathcal{K}_2(t,0)  +2\mathcal{K}_1^2\mathcal{K}_2(0,0) +  \notag \\ & +4\mathcal{K}_1 \mathcal{K}_3(0,t,0)+2\mathcal{K}^2_2(t,0) \ , \label{OTOCfreeprob} 
\end{align} 
where we recall that $\mathcal{K}_n$ are the ETH simple loops given by sums over different indices as in Eq.~\eqref{freeceth}. In particular, one has $\mathcal K_1 = \kappa_1$. 
In Fig.~\ref{fullETHOTOC}, we plot the exact result \eqref{OTOCexact} and the ETH prediction \eqref{OTOCfreeprob} for $N=6000$, which shows an excellent agreement, confirming the validity of full ETH in the chaotic regime.

We note that while $\mathcal{K}_2(t, 0)$ decays rapidly to zero, the dynamics of the OTOC are dictated mostly by high-order connected correlation functions and, in particular, by the fourth-order free cumulant, namely
\begin{equation}
    \langle \hat A(t) \hat A \hat A(t) \hat A\rangle \simeq C + \mathcal{K}_4(t, 0, t, 0)\ .
\end{equation}
This also shows that the late-time behaviour of the OTOC is encoded in $\mathcal K_{4}(t, 0, t, 0)$; as we saw above, this feature is generic and applies to arbitrary $2n$-OTOCs. 

At long times, the OTOC approaches a plateau given only by the time-independent contributions since the time-dependent free cumulants decrease to zero, displaying freeness. The value of the plateau $C=\mathcal{K}^4_1+ 2\mathcal{K}_1^2\mathcal{K}_2(0,0)$ corresponds to the one predicted by Free Probability for free-variables in Eq.~\eqref{fac} for $\hat{B}=\hat{A}$, i.e. 
\begin{equation}
    C = \mathcal K_1^2 \langle \hat A^2\rangle\ + \mathcal K_1^2 \mathcal K_2 \ .
\end{equation}

\begin{figure}[t]
\centering
\includegraphics[width=0.7\linewidth]{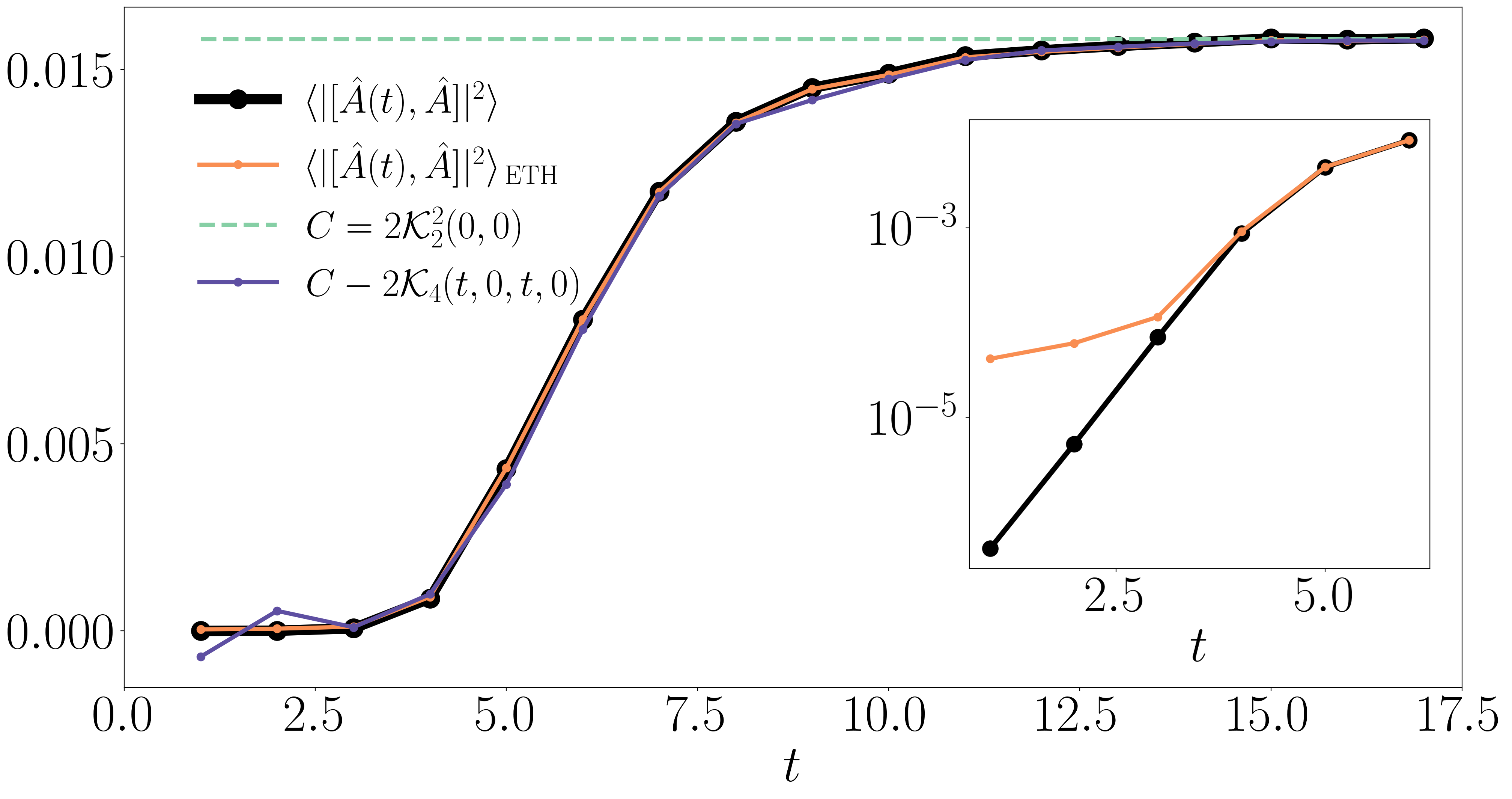} 
\caption{The exact dynamics of the square-commutator in Eq.~\eqref{SCCexact} (black line) is compared with the ETH approximation in Eq.~\eqref{decompSCC} (orange line). The dashed green line indicates the time-independent plateau, while the violet line shows the contribution due to the ETH-free cumulant $\mathcal{K}_4$. 
In the inset, we show the same data in a logarithmic scale at short times. The observable chosen is $\hat A$ in Eq.~\eqref{eq:obs}, for the kicked top \eqref{kickedtopHam} in the chaotic case $\gamma=6$, for $N=6000$. }
\label{SCCnumerical}
\end{figure}

\paragraph*{Square-commutator and ETH free cumulants}

We now show that free cumulants play a crucial role also in describing the dynamics of the so-called square-commutator, defined as
\begin{align}
\langle |[\hat{A}(t), \hat{A}] |^2 \rangle = - 2 \langle \hat{A}(t)\hat{A}\hat{A}(t)\hat{A} \rangle + 2 \langle \hat{A}^2(t)\hat{A}^2 \rangle\ ,
\label{SCCexact}
\end{align}
which contains, in turn, out-of-time order correlators. In systems with a well-defined classical chaotic limit, this object grows as  \cite{larkin1969quasiclassical, KitaTalk}
\begin{align}
\left \langle |[\hat{A}(t), \hat{A}] |^2 \right \rangle \sim \hbar_{\mathrm{eff}}^2 e^{2\tilde{\lambda} t},
\label{ExpGrowthOTOC}
\end{align}
until the Ehrenfest time (also called scrambling time)
\begin{equation}
    t_{\rm Ehr} \sim \log \hbar_{\mathrm{eff}}^{-1} / \tilde \lambda 
\label{eq:Ehrenfest}
\end{equation}
in the limit $h_{\rm eff}\to 0$. Beyond this time, quantum effects become significant, causing the square-commutator to saturate. The rate \(\tilde{\lambda}\) is often referred to as the quantum Lyapunov exponent \cite{garcia2022out, maldacena2016bound, hosur2016chaos, xu2022scrambling}.

We now explore how well the ETH predictions apply to Eq.~\eqref{SCCexact}. 
In the chaotic regime the square-commutator can be decomposed in terms of ETH-free cumulants as follows
\begin{align}
\langle |[\hat{A}(t), \hat{A}] |^2 \rangle_{\mathrm{ETH}} =& -2\mathcal{K}_4(t,0,t,0) + 2\mathcal{K}_4(t,t,0,0) - 2\mathcal{K}^2_2(t,0) + 2\mathcal{K}^2_2(0,0)\ ,
\label{decompSCC}
\end{align}
where the OTOC in Eq.~\eqref{SCCexact} is expanded as in Eq.~\eqref{OTOCfreeprob}, and the time-ordered correlator is factorized as\footnote{This decomposition highlights how certain ETH diagrams (see App.~\ref{app_fETH}), initially distinct between the two correlators before applying ETH, factorize in the same way for large system sizes. They thus cancel out at the leading order and give only subleading contributions to Eq.~\eqref{decompSCC}.}
\begin{align}
\langle \hat{A}^2(t)\hat{A}^2 \rangle_{\mathrm{ETH}} &= \mathcal{K}_4(t,t,0,0) + 
\mathcal{K}_1^4 + 4\mathcal{K}_1^2 \mathcal{K}_2(t,0) + 2\mathcal{K}_1^2\mathcal{K}_2(0,0) + \notag \\
&\quad+ 4\mathcal{K}_1 \mathcal{K}_3(0,t,0) + \mathcal{K}^2_2(0,0) + \mathcal{K}^2_2(t,0)\ .
\end{align}

In Fig.~\ref{SCCnumerical}, we compare the numerical results of the exact square-commutator with its ETH-free cumulants decomposition from Eq.~\eqref{decompSCC}, which shows very good agreement. By examining the individual free cumulants, a clear hierarchy emerges in their contributions to the square-commutator dynamics:
i) \(\mathcal{K}^2_2(0,0)\) determines the saturation value, being the only time-independent free cumulant, which serves as yet another manifestation of freeness; 
ii) \(\mathcal{K}^2_2(t,0)\) and \(\mathcal{K}_4(t,t,0,0)\) decay rapidly and fluctuate around zero;
iii) \(\mathcal{K}_4(t,0,t,0)\) drives much of the distinguishing dynamics of the square-commutator, particularly the evolution towards the plateau. Since this term encodes correlations among different matrix elements, the results confirm the idea of ascribing the peculiar dynamics of the square-commutator to the ETH correlations (see, e.g., Ref.~\cite{foini2019eigenstate}).


However, we note that the ETH alone is insufficient to fully capture the behaviour of the square-commutator at \emph{very early times}, as shown in the inset of Fig.~\ref{SCCnumerical}. While the square-commutator initially grows as \(\hbar_{\mathrm{eff}}^2 \sim D^{-2}\), the ETH decomposition is valid only at \(\mathcal{O}(D^{-1})\), leading to a parametric discrepancy between the two curves in the early-time regime. Therefore, for this class of models, the early-time exponential growth is not solely captured by \(\mathcal{K}_4(t,0,t,0)\), but rather involves subleading terms from both crossing loops and the decomposition of noncrossing terms. This calls for a better understanding of the subleading contributions and higher-order correction to the full ETH.

In conclusion, even if the ETH-free cumulants fail to reproduce the earliest dynamics, the numerical data show that the behaviour of the square-commutator is predominantly governed by the \(\mathcal{K}_4(t,0,t,0)\) cumulant, stemming from the OTOC.

\bigskip

\section{Time scale for the onset of freeness}
\label{sec_res}

In Section~\ref{sec_OTOCfree} we demonstrated that both $2n$-OTOCs and $2n$-free cumulants exhibit exponential decay in reaching long-time freeness, in the kicked-top model, meaning
\begin{equation}
\langle (\hat A_0(t) \hat A_0)^n \rangle \sim \text{exp}\left(- \frac{n}{\tau_n} t \right) \qquad \text{for } t \gg 1 \ .
\label{eq:2n_otoc_decay}
\end{equation}
The quantities $\tau_n$ quantify how quickly freeness is achieved. 
Building on these findings, we develop a large deviation framework (Sec.~\ref{sec_LD}) to understand this time scale, followed by a numerical analysis (Sec.~\ref{sec_num}) in the context of the kicked-top model.

\subsection{Large Deviation Theory for Freeness}
\label{sec_LD}

We consider generic $2n$-OTOCs of traceless observables that decay exponentially as
\begin{equation}
\langle (\hat A_0(t) \hat A_0)^n \rangle \sim e^{- R_n t} \qquad \text{for } t \gg 1 \ ,
\label{eq:2n_otoc_decay_}
\end{equation}
where for illustration purposes, we used $R_n=n/\tau_n$.
The $2n$-OTOC can be seen as moments of order $n$ of the spectrum of the operator $\hat A_0(t) \hat A_0$: we call $g_i(t)$ its complex eigenvalues\footnote{Note that the eigenvalues of  $\hat A_0(t) \hat A_0$  come in conjugate pairs, ensuring that the trace of its moments is a real number. This property follows from the fact that $\hat A_0(t) \hat A_0$ is a pseudo-hermitian operator, i.e. it obeys $(A_0(t)A_0)^{\dagger}= \eta ( A_0(t)  A_0) \eta^{-1}$ with $\eta= A_0$.
} and we assume that the modulus $|{g_i(t)}|$ has the time-dependent part given by $e^{- \rho_i(t) t }$. 
Specifically:
\begin{align}
g_i(t) = G_i  e^{- \rho_i(t) t  } e^{i \theta_i(t) }    \ .
\label{spectrum_gi}
\end{align}
In a generic system, the decay rates $\rho_i(t)$ are distributed and fluctuate over time. Therefore, probing their higher moments through $2n$-OTOCs provides insight into their dynamical distribution. \\

To describe their temporal fluctuations, we take inspiration from studies in multifractality, turbulence, and chaos \cite{benzi1984multifractal,paladin1987anomalous, paladin1986intermittency, crisanti1988lyapunov, evers200fluctuations, mace2019multifractal,sierant2022universal}.
After defining the probability distribution of the decay rates as $P(\rho_t)=\frac{1}{D}\sum^D_i \delta \left(\rho_t - \rho_i(t) \right)$, we \emph{assume the existence of a typical value $\rho_{\rm typ}$ and a Large Deviation form at long-times} for it:
\begin{equation}
    P(\rho_t)  \propto e^{\Gamma(\rho_t) t} \qquad \text{for } t \gg 1\ ,
\label{Plambda}
\end{equation}
where $\Gamma(\rho_t)$ is analogous to the Cramér function in large deviation theory, it is concave and has its maximum at the typical value where it also vanishes, i.e., $\Gamma(\rho_{\rm typ}) = 0$ and $\Gamma'(\rho_t) |_{\rho_{\rm typ}} = 0$.
We now demonstrate that assuming a large deviation principle for the spectrum of this operator yields specific properties for $\tau_{n}$.

Indeed, from Eq.~\eqref{Plambda} we can derive the exponential time-decay: 
\begin{align}
    \langle (\hat A_0(t) \hat A_0)^{n}\rangle & =  \frac{1}{D} \sum_i g^n_i = \frac{1}{D} \sum_i G^n_ie^{-n\rho_i(t) t} e^{i n\theta_i(t) } \notag \\  \propto \int^{\rho_\text{max}}_{\rho_\text{min}}d\rho &\,\, e^{ -\left( n\rho - \Gamma(\rho)\right)t}  \approx \displaystyle e^{-\min_{\rho}\left( n\rho - \Gamma(\rho) \right)t} \, \label{eq:comparison} , 
\end{align}
where the integral on the right-end side is solved using that, for $t\gg 1$, the dominant contribution comes from the inferior stationary point of the exponent function.
Comparing Eq.~\eqref{eq:comparison} with Eq.~\eqref{eq:2n_otoc_decay_} we get that 
\begin{equation}
    R_n = \min_{\rho}\left( n\rho - \Gamma(\rho) \right) 
\end{equation} 
is a Legendre transform of $\Gamma(\rho)$; as a consequence, since $\Gamma$ is a concave function of $\rho$, $R_{n}$ is also a concave function of $n$. Not only, $R_{n}/n$ decreases monotonically with $n$, leading to the inequality
\begin{align}
    R_{n} \le n \rho_{\rm typ} \ ,
\end{align}
where the typical value $\rho_{\rm typ}$ is given by
\begin{equation}
\label{rho_typ}
    \rho_{\rm typ} = \lim_{n \to 0} \frac{R_{n}}{n} = \frac{\partial R_{n}}{\partial n}\Big|_{n=0}\ .
\end{equation}


 Coming back to the times $\tau_{n} = n/R_{n}$,
it follows that they \emph{are monotonically increasing with $n$}, and, setting  $\tau_{\rm typ} = \rho_{\rm typ}^{-1}$, they satisfy
\begin{equation}
    \label{tau_n}
    \tau_{n} \geq \tau_{\rm typ}\ .
\end{equation}

Let us discuss some physical consequences.
The decay rate of the $2n$-free cumulants (or equivalently, the $2n$-OTOCs of traceless observables) quantifies how quickly freeness emerges and can be viewed as a probe for the temporal fluctuations in the spectrum of time-dependent observables. The freeness time-scale $\tau_n$, defined in Eq.~\eqref{defTauN}, capture the average decay of these moments and provide a meaningful description of the dynamical fluctuations, incorporating finite-time properties.\\
In the limit where fluctuations are absent\footnote{This holds if the probability distribution in Eq.~\eqref{Plambda} is $P(\rho) \propto \delta(\rho-\rho_{\rm typ})$.}, we find $\tau_n = \tau_{\rm typ}$. Analogous to studies of fractal dimensions in chaotic systems, systems exhibiting this property can be referred to as \emph{monofractal}. However, in general, we shall expect that the freeness time-scale $\tau_n$ will depend on $n$, implying a \emph{multifractal} structure in the approach to freeness.

\begin{figure}[t]
\centering
\includegraphics[width=0.7\textwidth]{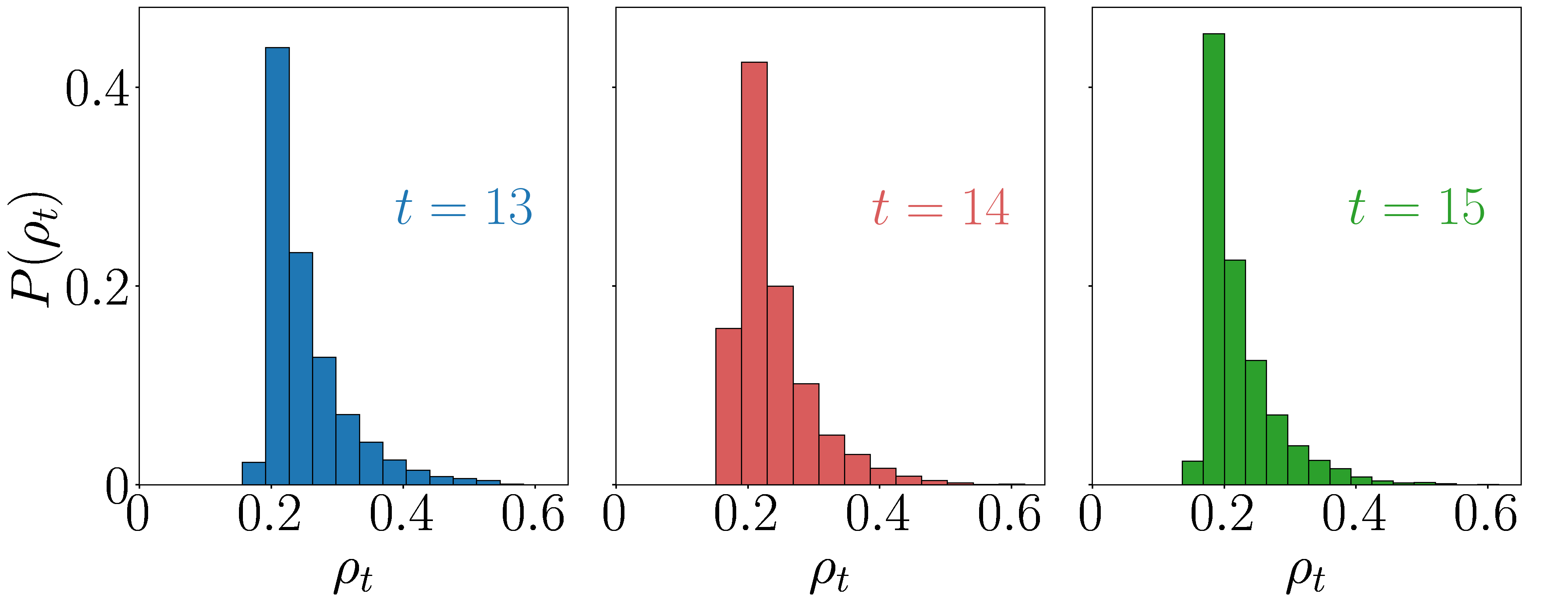}
\caption{Probability distribution $P(\rho_t)$ for the decay rates evaluated as $-\log{|g_i(t)|}/t$, for $t=13,14,15$, from the spectrum of $\hat A_0(t) \hat A_0$ (see Eq.~\eqref{spectrum_gi}). The traceless observable $\hat A_0$ is obtained from $\hat A$ in Eq.~\eqref{eq:obs}, for the kicked top \eqref{kickedtopHam} in the chaotic case $\gamma=6$, for $N=6000$.}
\label{LDassumtpions}
\end{figure}

\subsection{Freeness multifractality in the Kicked Top}
\label{sec_num}

Let us now turn to the numerical study of the onset of freeness in the chaotic regime of the kicked top. 

We have tested numerically that the large deviation ansatz in Eq.~\eqref{Plambda} is satisfied. In Fig.~\ref{LDassumtpions} we show the probability distribution $P(\rho_t)$ for the decay rates $\rho_i(t)$ defined in Eq.~\eqref{spectrum_gi}, evaluated as $-\log{|g_i(t)|}/t$, for $t=13,14,15$. 
The distribution is compatible with the large deviation ansatz in Eq.~\eqref{Plambda}: it gives rise to a concave exponent function that does not change in time and it is centered around a specific time-independent value $\rho_{\text{typ}}$.

\begin{figure*}[t]
\includegraphics[width=1\textwidth]{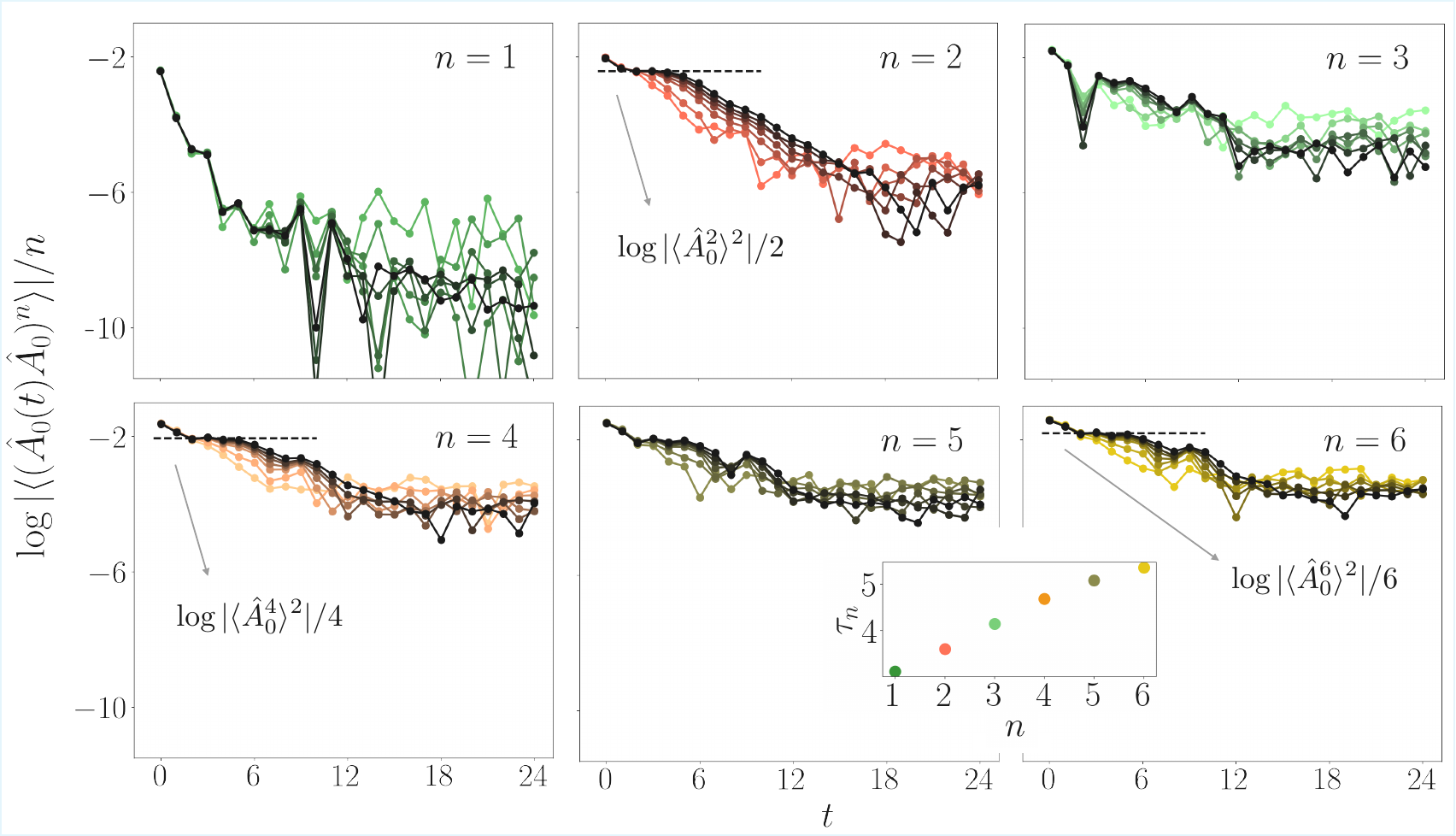}
\caption{The quantity $\displaystyle  \log | \langle (\hat A_0(t) \hat A_0)^n \rangle | / n$ as a function of time for different $n$,  from left to right $n=1,2,3,4$. In each figure, the system size is increased as $N \in [300,600,1400,3000,4000,6000]$, from light to dark colors. In the inset the freeness time-scale $\tau_n$ as a function of $n$, obtained through linear fits of the curves in an intermediate regime, for $N=6000$.
The traceless observable $\hat A_0$ is obtained from $\hat A$ in Eq.~\eqref{eq:obs}, for the kicked top \eqref{kickedtopHam} in the chaotic case $\gamma=6$.}
\label{CumulantsD}
\end{figure*}

We now report the study of the freeness time-scale $\tau_n$ as defined from Eq.~\eqref{defTauN}. 
In Fig.~\ref{CumulantsD} we plot the rescaled $2n$-OTOC
\begin{equation}
\notag
    \frac{1}{n} \log \left| \langle(\hat A_0(t) \hat A_0 )^n \rangle \right | \,\, 
\end{equation}
for $n=1,2,3,4$, from left to right.
The plot shows that the $2n$-OTOCs decay exponentially fast in time, according to Eq.~\eqref{eq:2n_otoc_decay} (linear decay in this scale). 
The data suggest that freeness is slower for increasing order $n$, as predicted by our large deviation theory in the case of multifractal behaviour.

To qualitatively address this issue, we evaluate the freeness time-scale $\tau_n$ by performing fits on the curves shown in Fig.~\ref{CumulantsD}, for $N=6000$. The fits are conducted within an intermediate regime, which is selected by omitting the initial time steps and extending up to the point where the plateau is reached. Due to the noise in the data, the fitting window was chosen 
individually for each $n$, after making an interpolation of the curves in time in order to have more points.
We show $\tau_n$ as a function of $n$ in the inset of Fig.~\ref{CumulantsD} (right-end panel), up to $n=6$ (the dynamics for $n=5, 6$ is not shown). The data shows that $\tau_n$ is a monotonic increasing function of $n$, thus confirming the expected behaviour from the large deviation theory presented above, cf. Eq.~\eqref{tau_n}. The extracted values of $\tau_n$ should be regarded as indicative: the analysis is mainly intended to highlight the qualitative trend rather than to quantify the functional dependence on $n$.

As discussed in Sec.~\ref{sec_tran} above, at longer times, the $2n$-OTOCs display fluctuations around a plateau $\mathcal O(D^{-1})$ indicating freeness.\\

From Fig.~\ref{CumulantsD}, we also highlight a qualitative difference between even and odd powers.
In the case of \emph{even} $n$, the $2n$-OTOC 
display a \textit{time shift}, which grows logarithmically with the system size. 

This is shown in Fig.~\ref{FreenessRate}, where we plot the time $t_\epsilon$ at which each $2n$-OTOC reaches an $\epsilon$ value, as a way to quantify the shift. For definiteness, we fix $\epsilon=-2.55$ and calculate $t_\epsilon$ as the first interpolated time for which $ \log \left| \langle (\hat A_0(t) \hat A_0)^n \rangle \right|/n \leq \epsilon$, for $n=1,2,3,4$. The data show a clear logarithmic increase of $t_\epsilon$ for even $n$, while the time is constant for odd $n$.
Furthermore, we also note that the time shift depends on the amount of chaos present in the system. We calculate the time shift also for $\gamma=4$ and $n=2$ corresponding to the mixed phase-space region discussed, see e.g. Fig.~\ref{fig_1}(left) in Sec.~\ref{sec_tran}. In this case, the time shift $t_\epsilon$ is bigger than the corresponding one for $\gamma=6$, indicating a smaller shift for more chaotic systems.

In the case of models with a semiclassical chaotic limit $\hbar_{\rm eff}\to 0$, the explanation for this effect can be found in the early-time dynamics of the $2n$-OTOCs. 
This behaviour has been studied for single-body quantum chaos \cite{garcia2018chaos, notenson2023classical}, as well as for the Sachdev-Ye-Kitaev (SYK) model \cite{kobrin2021many}, for $n=2$. In this case, the OTOC is known to have a Lyapunov regime at short intermediate times, meaning it is governed by the exponential growth of the square-commutator (cf. Eq.~\eqref{ExpGrowthOTOC}). 
In particular,
\begin{subequations}
\label{eq:OTOCSCC}
\begin{equation}
\langle \hat A(t)\hat A\hat A(t) \hat A\rangle \sim \text{const.} - e^{2\tilde{\lambda}(t-t_{\rm Ehr})}\, 
\label{eq:OTOCSCC_2}
\end{equation}
for $1\ll t\ll t_{\rm Ehr}$, with the Ehrenfest time defined in Eq.~\eqref{eq:Ehrenfest}. After this time the square-commutator saturates. 
\begin{figure}[t]
\centering
\includegraphics[width=0.7\textwidth]{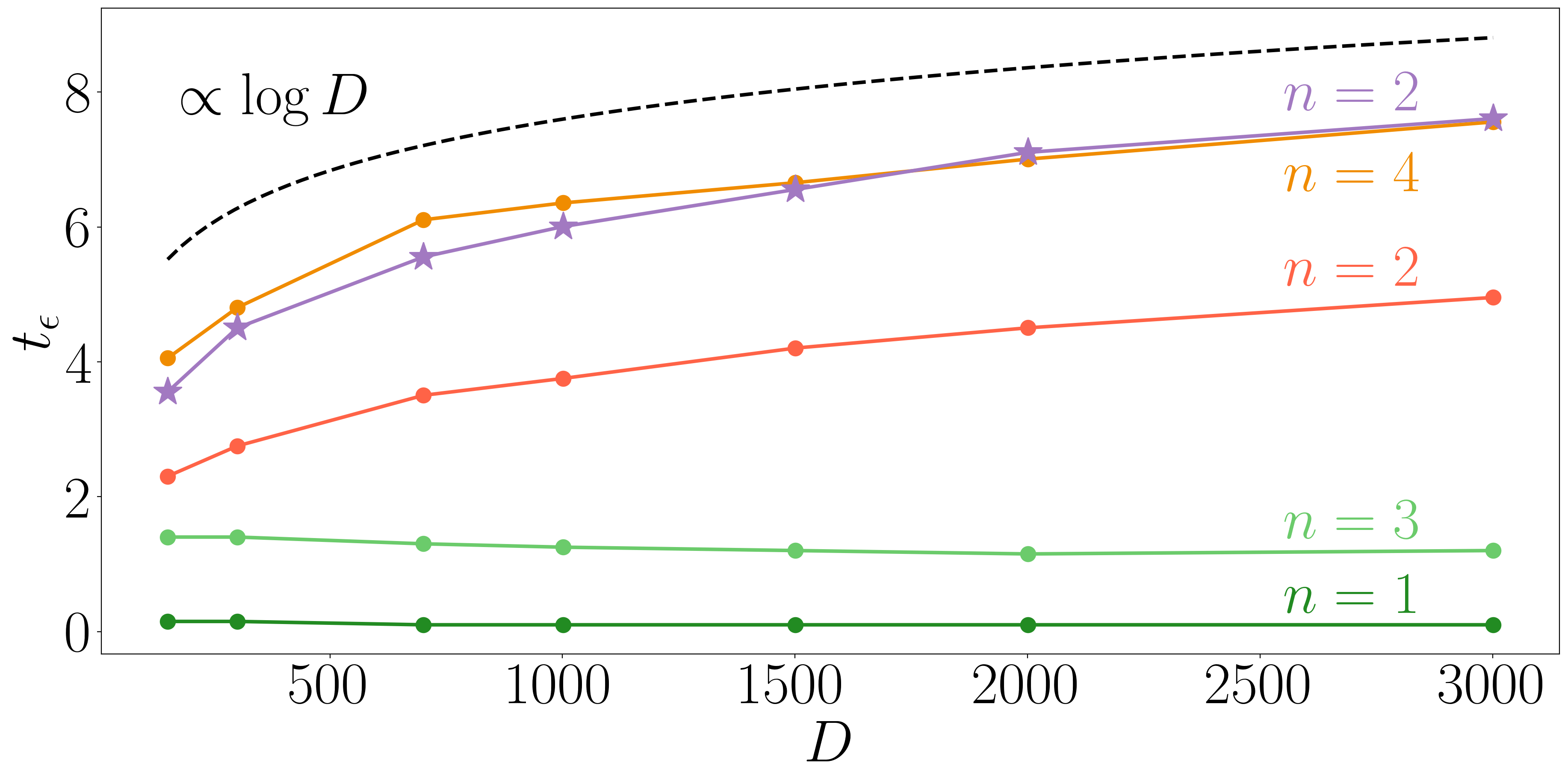}
\caption{$t_{\epsilon}$ defined as the first interpolated time for which $\displaystyle \frac{1}{n} \log  | \left\langle (\hat A_0(t) \hat A_0)^n \right \rangle  | < \epsilon=-2.55$ as a function of the system size, for different $n$. 
Points markers indicate the chaotic case $\gamma=6$; the colors refer to the associated curves in Fig.~\ref{CumulantsD}. The purple-star curve, instead, indicates the case $\gamma=4$, referring to Fig.~\ref{fig_1}. The traceless observable $\hat A_0$ is obtained from $\hat A$ in Eq.~\eqref{eq:obs}, for the kicked top \eqref{kickedtopHam}.}
\label{FreenessRate}
\end{figure} 

For higher and even $n=2m$, a similar behaviour to the one of Eq.~\eqref{eq:OTOCSCC_2} is conjectured to be valid at short intermediate times \cite{Pappalardi2023quantum}, i.e.
\begin{equation}
\langle (\hat A(t)\hat A )^{2m}\rangle \sim \text{const.} - e^{L_{2m}\left(t-t^{(2m)}_{\rm Ehr}\right)}\, 
\label{eq:OTOCSCC_n}
\end{equation}
\end{subequations}
for $1\ll t\ll t^{(2m)}_{\rm Ehr}$,
where $L_{2m}$ are the rates accounted by the so-called generalized Lyapunov exponents, which stem from the powers of the square-commutator ${\langle| [\hat A(t), \hat A(0)]|^{2m}\rangle \sim \hbar_{\rm eff}^{2m} e^{L_{2m}t}}$.
One retrieves the square commutator for $m=1$ ($n=2$). This behaviour is valid until the generalized Ehrenfest time ${t^{(2m)}_{\rm Ehr}\sim \log \hbar_{\rm eff}^{-2m}/L_{2m}}$, which is assumed to be increasing with $m$, see also Refs.\cite{trunin2023quantum, trunin2023refined}. After this time scale, the higher powers of the square-commutator saturate. In the case of odd powers, this effect is not present, indeed, the odd powers of the commutator usually do not have exponential growth due to a cancellation of signs. The Lyapunov exponents obey a generalized bound to chaos, i.e. $L_{2m} \leq 2 \pi m/\beta \hbar$ \cite{tsuji2018bound, Pappalardi2023quantum}. However, since the model under scrutiny does not conserve energy, i.e. it is at infinite temperature, these bounds do not play a role for our discussions. 

The behaviour described in Eq.~\eqref{eq:OTOCSCC_n} holds up to $t^{(2m)}_{\rm Ehr}$, which marks the saturation of higher powers of the square-commutator and correspondingly the initial point after which the $4m$-OTOC exhibits the exponential decay. Since for this model $\hbar_{\rm eff}^{-1}\sim D$, this leads to \begin{equation}
    t^{(2m)}_{\rm shift} \sim t^{(2m)}_{\rm Ehr} \sim \frac {2m}{L_{2m}} \log D\ .
\end{equation}
In summary, the observed time shift is explained through the Ehrenfest time: the exponential decay of the $4m$-OTOCs starts later for bigger $D$, and it is inversely proportional to the amount of chaos, thus explaining the results in Fig.~\ref{FreenessRate}. 
As already noted in Sec.~\ref{sec_tran}, we observe that, also at the level of $2n$-OTOCs, classical independence emerges before the onset of freeness. When the observables become classically independent, the $2n$-OTOCs factorize in a simple way, $\langle (\hat A_0(t) \hat A_0)^n \rangle = \langle \hat A_0^n \rangle^2$, as if $\hat A_0$ and $\hat A_0(t)$ were commuting operators. This corresponds to the plateau value reached by the $4n$-OTOCs at short times, up to the generalized Ehrenfest time, and is represented by the dashed line in Fig.~\ref{CumulantsD}. For odd powers, $\langle \hat  A_0^n \rangle = 0$ for the specific choice of observable considered here (e.g., due to its symmetric spectrum), and therefore no plateau is observed in those cases. As discussed above, classical independence and freeness cannot coexist and instead appear sequentially in time, as the semiclassical system of finite dimension transitions from classical behaviour to a fully quantum regime.

\section{Conclusions and Discussion}
\label{sec_disc}

In this work, we explored the emergence of long-time freeness induced by chaotic dynamics in the kicked top, a prototypical model of quantum chaos. 
We established that long-time freeness is reached exponentially fast in the chaotic regime.
This led us to introduce a large deviation theory for the approach to freeness and the associated time scale.

Our results indicate that Free Probability could have a wider range of applicability in the study of many-body quantum systems. 
We conclude with a discussion of potential future directions stemming from this work, which would be valuable to explore.

\begin{itemize}
    \item \emph{Large deviation theory of freeness.} Guided by approaches in turbulence, chaos, and eigenstate delocalization, in this work, we introduced a large deviation theory of freeness, which shall apply generically to exponentially decaying $2n$-OTOCs. The associated freeness time-scale $\tau_n$ describe the time-dependent fluctuations of the spectrum of time-evolving observables and indicate a multifractal behaviour if $\tau_n>\tau_{\rm typ}$ dependent on $n$, and a form of monofractality if $\tau_n=\tau_{\rm typ}$ for every $n$.
In chaotic single-body quantum systems with classical limits, the exponential decay of the $4$-OTOC was interpreted from the point of view of mixing and associated with Ruelle-Pollicot resonances by Refs.\cite{garcia2018chaos, notenson2023classical}, see also Ref.~\cite{polchinski2015chaos}.   Within this framework, it would be intriguing to explore whether the mono/multifractal behaviours in the large deviation theory of freeness can be interpreted in terms of these resonances.

    Although our analysis stems from chaos in single-body semiclassical systems, it should extend to any context where freeness is achieved exponentially, including many-body systems. For example, recent findings report exponential decay in free cumulants of order $n$ within dual-unitary circuits \cite{chen2024free}, suggesting promising directions for future exploration of the freeness time-scale in many-body settings.

    Another potential future direction would amount to exploring the effects of locality in the large deviation theory of freeness, leading to questions also about the (potentially $n$-dependent) butterfly velocity and its behaviour in local systems, such as clean or random circuits.

Finally, clarifying the precise relation between the different timescales associated with the emergence of freeness in chaotic quantum dynamics, recently investigated from complementary perspectives in Refs.~\cite{fritzsch2025free, vardhan2025free,fritzsch2025free2,jahnke2025free}, represents an interesting open direction for future work.

    \item {\emph{Beyond full ETH.}}
    Throughout this work, we have focused on expectation values, whose leading behaviour is well described by ETH-free cumulants. However, ETH predictions are accurate only at the leading order and do not fully account for subleading effects, which appear at order $\mathcal O(D^{-1})$.

    For instance, we have shown that the leading-order ETH predictions fail to capture the early-time behaviour of the square-commutator (see Sec.~\ref{sec_ETH}) or the late-time fluctuations of the free cumulants (see Sec.~\ref{sec_tran}). This highlights the need to extend ETH to account for finite system size corrections.

Finally, we explored the development of freeness and free-cumulant dynamics up to their \( \mathcal{O}(1/D) \) saturation level. At even longer time scales, signatures of level repulsion may appear. This effect is known at the level of the two-point function \cite{cotler2017black, delacretaz2020heavy, altland2021from, yoshimura2023operator, bouverot2024random}, and it is also expected at higher orders \cite{cotler2017chaos, cotler2020spectral}. While this is consistent with asymptotic long-time freeness,  it invites further investigation into how higher-order spectral fluctuations manifest in observable quantities.

\end{itemize}

\section*{Acknowledgements}
We thank X. Turkeshi, B. Pain, and J. Kudler-Flam for their insightful discussions, M. Buchhold, L. Foini and Y. Fyodorov for valuable feedback on our work, and J. Kurchan for collaboration on related projects. 
We acknowledge support by the Deutsche Forschungsgemeinschaft (DFG, German Research Foundation) under Germany’s Excellence Strategy - Cluster of Excellence Matter and Light for Quantum Computing (ML4Q) EXC 2004/1 -390534769, and DFG Collaborative Research Center (CRC) 183 Project No. 277101999 - project B02.\\
The data that support the findings of this article are openly available \cite{zenodo2024}.

\appendix

\section{Diagrammatic representation of the Full Eigenstate Thermalization Hypothesis}
\label{app_fETH}

\begin{figure*}[t]
\includegraphics[width=1\textwidth]{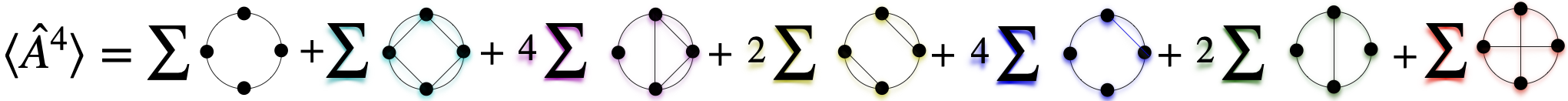}\\
\includegraphics[width=0.7\textwidth]{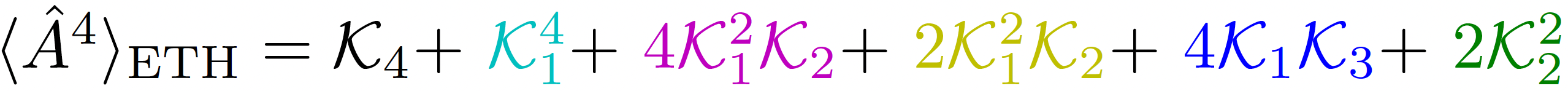}
\caption{Four-time correlation function. In the first row, pictorial representation: the sum in Eq.~\eqref{correlation4} for $n=4$ has been split in sums over all possible combinations of repeated indices or, in other words, in all the possible ETH diagrams with four dots; the result is made of sums of simple, non crossing and crossing loops.
In the second row, simplification, as ETH is applied: non crossing diagrams factorize in simple diagrams whose contribution is given by Eq.~\eqref{freeceth}, whereas the crossing diagram is negligible; thanks to these two conditions, $\mathcal{K}_n$ are identified as free cumulants, for large $D$. Indeed, while the decomposition in the first row is exact, the second one is true at the first order in the system size $D$.  
One diagram and its factorization are indicated with the same color. We mention that some loops have cyclic permutations, indicated by the numbers in front of the sums, which in general lead to free cumulants with different time dependences (here we considered equal-time, for convenience).}
\label{pictorial4corr}
\end{figure*}

The full Eigenstate Thermalization Hypothesis ansatz in Eq.~\eqref{full_ETH} can be represented in a pictorial way via \emph{ETH diagrams}; the product of $n$ matrix elements is drawn as a loop with $n$ dots, where every dot indicates an index and every arc of the circle between two consecutive dots refers to the matrix element with the related indices. By convention, the factors of the product are written down clockwise along the loop, starting from the upper dot. In the presence of repeated indices, lines are drawn to connect dots with equal indices. One recognizes three different types of diagrams (simple loops, noncrossing, and crossing) based on the drawn lines. 
Some examples of ETH diagrams for $n=6$ are here represented:\\
\begin{center}
\includegraphics[width=.8\textwidth]{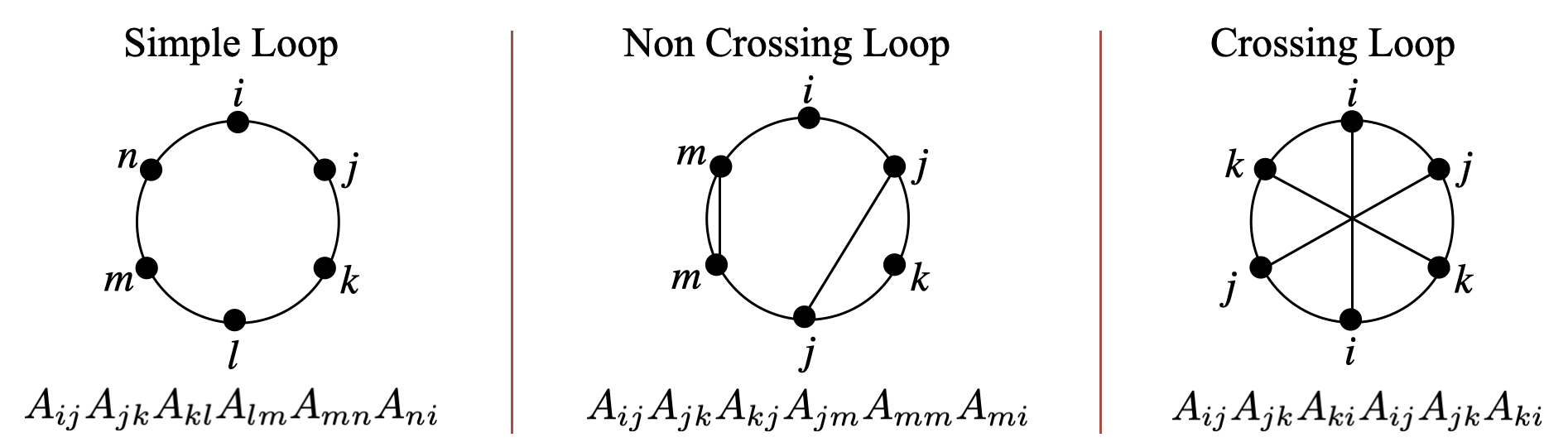}\end{center}

The full ETH ansatz permits to study equilibrium multi-time correlation functions, i.e. 
\begin{align}
\langle & \hat{A}(t_1) \dots  \hat{A}(t_n) \rangle = \frac{1}{D} \sum_{\alpha_1, \dots, \alpha_n}  e^{i \vec \omega \cdot \vec t} A_{\alpha_1 \alpha_2}\dots A_{\alpha_n \alpha_1}\ , 
\\ & \text{with}\quad \vec \omega=(\omega_{\alpha_1 \alpha_2}, \dots,\omega_{\alpha_{n} \alpha_1}) \ ,\quad \vec t=(t_1,\dots,t_n) \notag
\end{align}
We now illustrate how within ETH the building blocks of multi-time correlations are $\mathcal{K}_n(\vec t \,)$ as given in Eq.~\ref{freeceth}. \\
We start from the sum over all the indices in the correlation: this can be split in one sum with all different indices $(\sum_{\alpha_1 \ne \dots \ne \alpha_n})$ plus sums in which only one index is repeated $(\sum_{\alpha_1=\alpha_2 \ne \alpha_3 \ne \dots \ne \alpha_n}$ for instance), plus sums with two repeated indices and so on. 
In a pictorial fashion, the correlation function is drawn through sums of $n$-dots loops with all the possible contractions of lines: a sum over a $n$-dots simple loop plus sums over all possible crossing and noncrossing loops (see Fig.~\ref{pictorial4corr} for the case $n=4$). At this level, the ETH diagrams have been used only as bookkeeping; their usefulness arises when one applies the full ETH at the level of each sum:
\begin{itemize}
    \item the sum of a simple loop is given by
    \begin{equation}
    \mathcal{K}_n(\vec t \,) := \frac{1}{D} \sum_{\alpha_1\ne\dots \ne \alpha_n}  e^{i \vec \omega \cdot \vec t}  A_{\alpha_1 \alpha_2} \dots A_{\alpha_n \alpha_1} \ .
\end{equation} 
    According to Eq.~\eqref{eth1}, this quantity can be written as 
    $\mathcal{K}_n(\vec t\,)= \int d \vec{\omega}  e^{i \vec \omega \cdot \vec t} F^{(n)}(\vec \omega)$ which states that ETH smooth functions are linked to the Fourier Transform of $\mathcal{K}_n(\vec t\,)$. For instance,
    \begin{equation}
    \centering
    \includegraphics[width=0.85\textwidth]{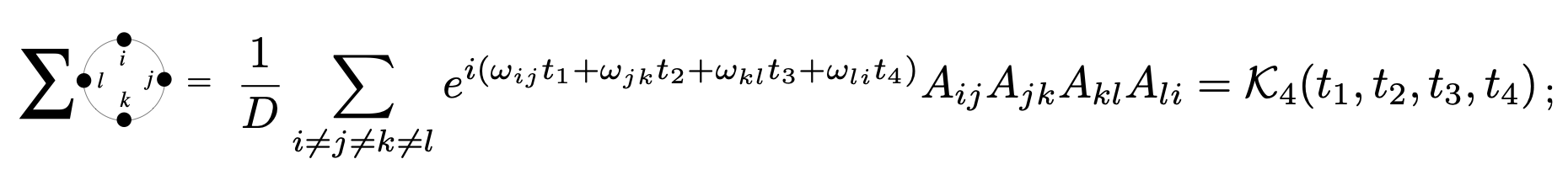}
    \end{equation}
    \item the sum of a noncrossing loop factorizes in the sums of simple loops with lower numbers of dots. For instance, according to Eq.~\eqref{eth2} applied on the product of matrix elements
    \begin{align}
    \includegraphics[width=.6\textwidth]{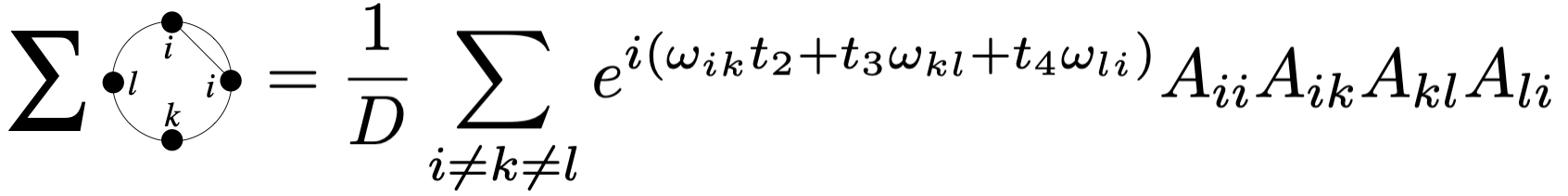} \notag\\ \includegraphics[width=.6\textwidth]{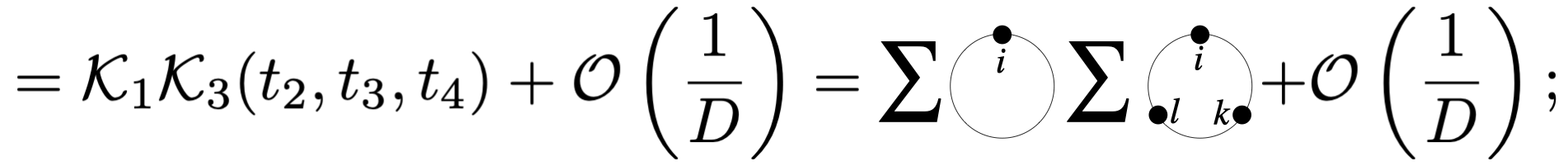}        
    \end{align}
  
    \item the sum of a crossing loop, instead, gives a negligible contribution at the first order in the system size. For instance, according to Eq.~\eqref{eth2} applied on the product of matrix elements 
\begin{equation}
\centering
    \includegraphics[width=.58\textwidth]{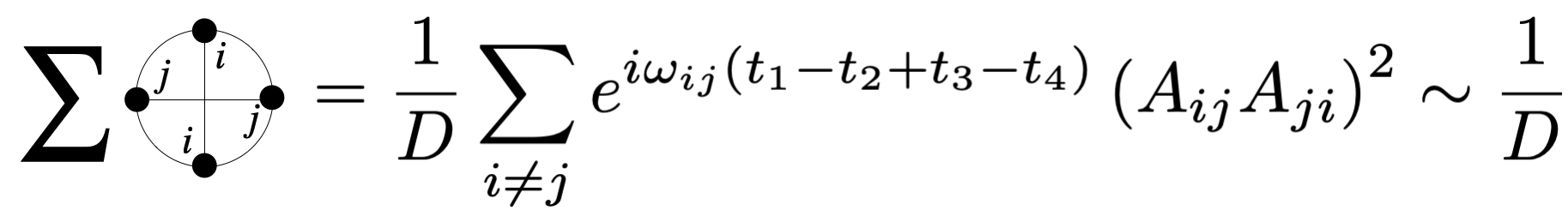}
\end{equation}    
    and can therefore be neglected.
\end{itemize}

Summarizing, the full Eigenstate Thermalization Hypothesis can be reformulated as the \emph{two ETH properties}:
\begin{enumerate}
    \item suppression of crossing diagrams, e.g.
\begin{center}
\vspace{-.2 cm}
    \includegraphics[width=.17\textwidth]{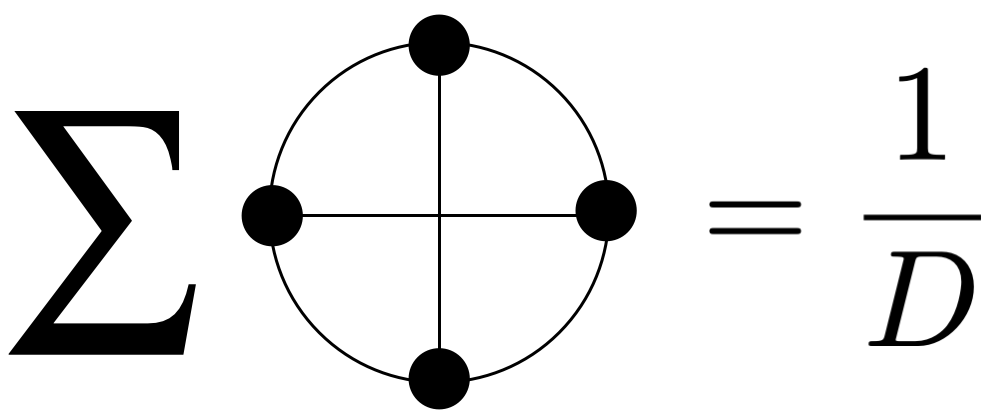}\, ; 
\end{center}
    \item factorization of noncrossing diagrams in simple loops, e.g. \,
\begin{center}
\vspace{-.1 cm}
    \includegraphics[width=.33\textwidth]{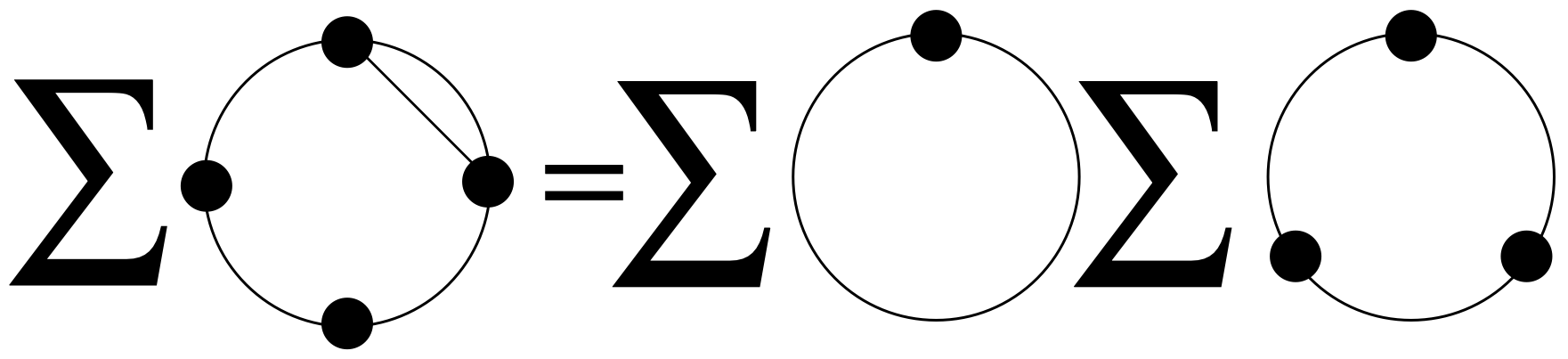} \, ,
\end{center}
\end{enumerate}
at the leading order in the system size. 

Hence, the simplified result for correlations calculated through ETH, for large system sizes, is a sum of products of $\mathcal{K}_n$ for different $n$ i.e. diagrams involving only sums over different indices; these building blocks of multi-time correlations are called plainly ``ETH simple loops'' in the text. The explicit formula for the case $n=4$ is reported in Fig.~\ref{pictorial4corr}.

\section{Derivations of long-time freeness}
\label{appB}
We review the proof of Eqs.\eqref{eq_infinity}-\eqref{eq_infinity_2} by keeping the finite size corrections. We focus on $\kappa_2(t)= \langle \hat A(t) \hat A\rangle-\braket{\hat A}^2$, which can be re-expressed as \cite{pappalardi2022eigenstate}
\begin{align}
    \begin{split}
          \kappa_2(t,0) & = \frac 1D\sum_{i \neq j }A_{ij}A_{ji} e^{i(E_i-E_j)t} 
          +\frac 1D \sum_i A_{ii}^2 - \left ( \frac 1D \sum_i A_{ii} \right )^2 =
          \\
          & = \frac 1D\sum_{i \neq j }A_{ij}A_{ji} e^{i(E_i-E_j)t} + \mathcal O(D^{-1})\ ,
    \end{split}
\end{align}
where from the first to the second line we have used that $\overline{A_{ii}^2} = \overline{A_{ii}}^2 + F^{(2)}(\omega=0)/D$. Taking the infinite-time limit we obtain
\begin{align}
\label{long-time}
    [\kappa_2]_\infty=\mathcal O(D^{-1})\ ,
\end{align}
using that $\displaystyle \lim_{T\to \infty} \frac 1T \int_{0}^T dt \sum_{i\neq j}e^{i(E_i-E_j)t}=0$. This is true due to the non-resonance condition.\\ 
We now evaluate the variances of such fluctuations:
\begin{align}
\label{long-time-2}
\begin{split}
    & [\kappa_2^2]_\infty - [\kappa_2]^2_\infty  = \lim_{T\to \infty} \frac 1T  \int_{0}^T dt\, \left ( \frac 1D\sum_{i \neq j }A_{ij}A_{ji} e^{i(E_i-E_j)t} \right)^2 + \mathcal O(D^{-2}) = 
    \\
    & = \frac 1{D^2}\sum_{i\neq j}  A_{ij}A_{ji}A_{ij}A_{ji}
    + \mathcal O(D^{-2}) = \mathcal O(D^{-2})   \ . 
\end{split}
\end{align}
From the first to the second line we have used the non-resonance condition which implies 
$\displaystyle \lim_{T\to \infty} \frac 1T \int_{0}^T dt\, e^{i(E_i-E_j)t+i(E_k-E_m)t} = \delta_{ij}\delta_{km} + \delta_{im}\delta_{jk}$. The last equality, instead, follows from the full ETH scaling $\overline{A_{ij}A_{ji}A_{ij}A_{ji}}\sim D^{-2}$. \\
The infinite-time average of Eqs.\eqref{long-time}-\eqref{long-time-2} can be generalized to higher $n$, leading to Eqs.\eqref{eq_infinity}-\eqref{eq_infinity_2} of the main text. This is done by means of the full ETH, and of the generalized non-resonance condition. 

\section{Slowest dynamics of the alternating free cumulants}
\label{app_C}
\begin{figure}[t]
\centering
\includegraphics[width=1\textwidth]{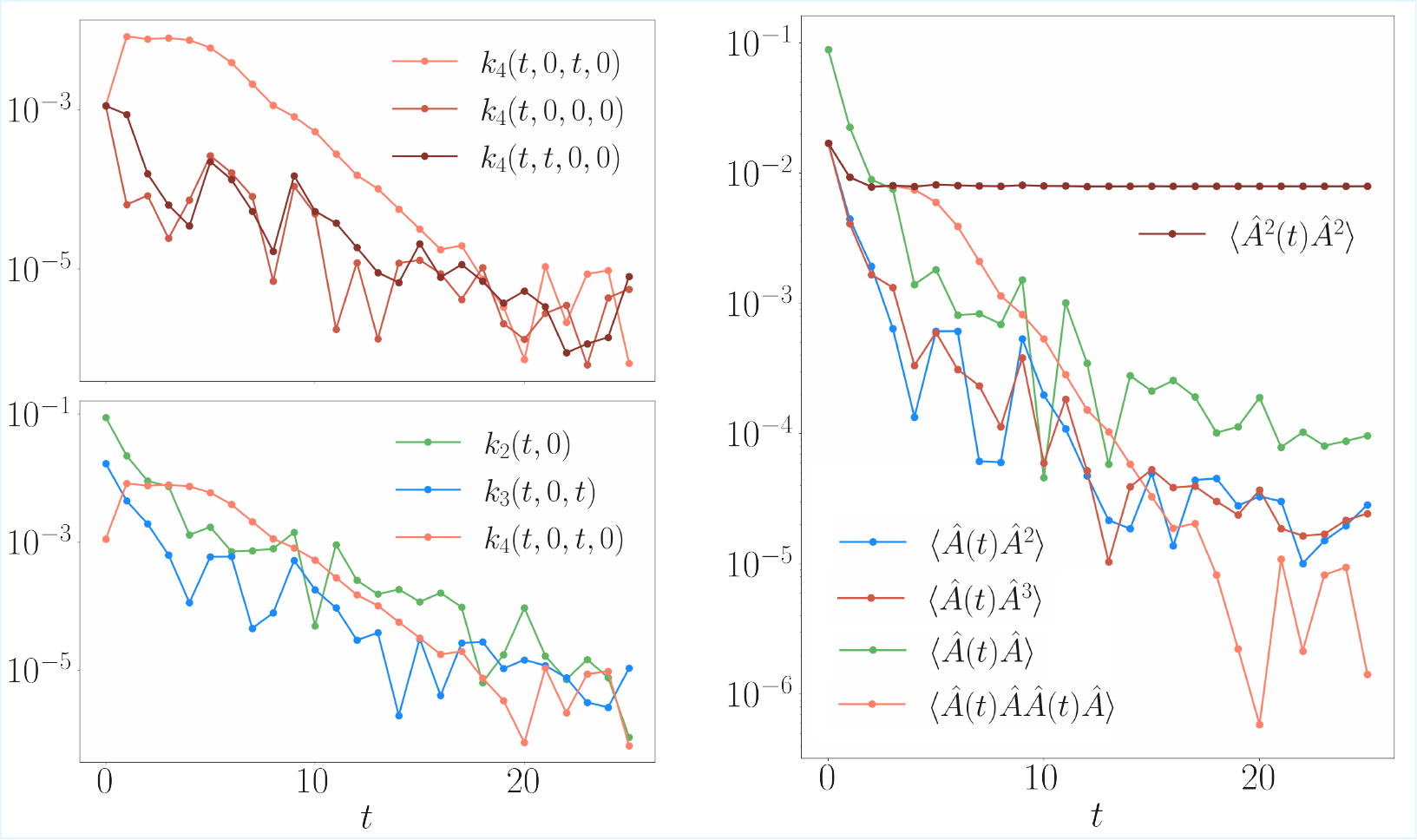}
\caption{Numerical evidence for the slowest decay of alternating free cumulants and moments among all the mixed ones, in the kicked top \eqref{kickedtopHam}, for the traceless observable $\hat A_0$ obtained from $\hat A$ in Eq.~\eqref{eq:obs}, in the chaotic regime $\gamma=6$, for $N=6000$.\\
(left) Top, among the fourth-order free cumulants, the alternating one is the last to decay. Bottom, $\kappa_3(t,0,t)$ exhibits a decay behaviour similar to that of $\kappa_2(t,0)$. \\
(right) Evolution in time of the mixed moments corresponding to the mixed free cumulants shown in the adjacent panel, for comparison.}
\label{fig_late}
\end{figure}

In the main text, we focus exclusively on alternating free cumulants, as we claim that they exhibit the slowest dynamics and are therefore the most suitable, among all mixed free cumulants, for defining a characteristic freeness time scale. The justification of this statement lies in the number of phase factors appearing in the different mixed free cumulants. To clarify this point, we consider two extreme cases and evaluate the free cumulants as sums over unequal indices, as in Eq.~\eqref{freeceth}.
\begin{itemize}
    \item In the case of complete separation of times, $\kappa_{2n}(t,\dots,t,0,\dots,0)$, when expressing this correlator in the energy basis, the $n$ temporal phase factors, being consecutive, combine into a single phase factor that depends only on the difference between the first and the last energy. When summing, the phase factors from different terms of the sum will become independent in a short time and cause the early decay of the free cumulant.
    \item In the alternating case $\kappa_{2n}(t,0,\dots,t,0)$, the $n$ temporal phase factors do not collapse into a single contribution; instead, each depends on the difference between consecutive energies. As a result, the phase factors in different terms of the sum remain correlated for longer times, requiring more time to dephase and cancel out, which leads to a later decay of the correlator.
\end{itemize}
Intermediate cases display decay times that lie between these two extremes, which explains why alternating free cumulants are the last to decay among all mixed ones. \\ One may also ask what happens for odd alternating free cumulants, for instance $\kappa_{2n+1}(t,0,\dots,t,0,t)$. Their decay behaviour is essentially the same as that of $\kappa_{2n}(t,0,\dots,t,0)$, although their magnitude is smaller. For this reason, we restricted our analysis to the even case, for which it is also more straightforward to formulate a large deviation theory, with the understanding that analogous arguments apply separately to the odd case. A similar picture holds for the moments: alternating $2n$-OTOCs decay slower than other mixed correlations, and even and odd cases exhibit comparable decay rates, e.g. $\langle(\hat A(t)\hat A)^n \hat A(t)\rangle$ decays similarly to $\langle(\hat A(t)\hat A)^n\rangle$. However, in contrast to free cumulants, not all mixed moments decay to zero; for instance, while $\kappa_4(t,t,0,0)$ vanishes at long times, the corresponding moment $\langle \hat A^2(t)\hat A^2\rangle$ approaches a constant value, namely $\langle\hat A^2\rangle^2$. \\
We support these considerations with numerical results in Fig.~\ref{fig_late}. We analyse numerically the dynamics of the mixed fourth-order free cumulants and compare them together. Specifically we consider $\kappa_4(t,0,t,0), \kappa_4(t,t,0,0)$ and $\kappa_4(t,0,0,0) (= \kappa_4(t,t,t,0))$. Among these free cumulants, the alternating one is the slowest to decay to zero. We also compare the third-order alternating free cumulant with the second and the fourth orders, in order to show that the decay rate of $\kappa_{2+1}(t,0,t)$ is essentially the one of  $\kappa_{2}(t,0)$. Finally, we plot all the respective mixed moments for comparison: while all the mixed free cumulants decay to zero due to freeness of the entries, this is not the case for the mixed moments, which instead reach the value predicted by Eq.~\eqref{eq_free_prod}. However, the relation between the dynamics of the different mixed moments is the same as for the free cumulants. In particular, $\langle \hat A(t) \hat A\rangle$, $\langle \hat A(t) \hat A^2\rangle$ and $\langle \hat A(t) \hat A^3\rangle$ exhibit essentially the same dynamics, and they vanish faster than $\langle \hat A(t) \hat A\hat A(t)\hat A\rangle$.

\bibliographystyle{quantum}
\bibliography{biblio}

\begin{thebibliography}{100}

\bibitem{delsarte1991spherical}
Philippe Delsarte, Jean-Marie Goethals, and Johan~Jacob Seidel.
\newblock ``Spherical codes and designs''.
\newblock In Geometry and Combinatorics.
\newblock \href{https://dx.doi.org/10.1007/BF03187604}{Pages 68--93}.
\newblock Elsevier~(1991).

\bibitem{dankert2005efficient}
Christoph Dankert.
\newblock ``Efficient simulation of random quantum states and
  operators''~(2005).
\newblock
  \href{http://arxiv.org/abs/quant-ph/0512217}{arXiv:quant-ph/0512217}.

\bibitem{dankert2009exact}
Christoph Dankert, Richard Cleve, Joseph Emerson, and Etera Livine.
\newblock ``Exact and approximate unitary 2-designs and their application to
  fidelity estimation''.
\newblock \href{https://dx.doi.org/10.1103/PhysRevA.80.012304}{Phys. Rev. A
  {\bf 80}, 012304}~(2009).

\bibitem{cotler2023emergent}
Jordan~S Cotler, Daniel~K Mark, Hsin-Yuan Huang, Felipe Hernandez, Joonhee
  Choi, Adam~L Shaw, Manuel Endres, and Soonwon Choi.
\newblock ``Emergent quantum state designs from individual many-body wave
  functions''.
\newblock \href{https://dx.doi.org/10.1103/PRXQuantum.4.010311}{PRX Quantum
  {\bf 4}, 010311}~(2023).

\bibitem{Choi_2023}
Joonhee Choi, Adam~L. Shaw, Ivaylo~S. Madjarov, Xin Xie, Ran Finkelstein,
  Jacob~P. Covey, Jordan~S. Cotler, Daniel~K. Mark, Hsin-Yuan Huang, Anant
  Kale, Hannes Pichler, Fernando G. S.~L. Brand{\~{a}}o, Soonwon Choi, and
  Manuel Endres.
\newblock ``Preparing random states and benchmarking with many-body quantum
  chaos''.
\newblock \href{https://dx.doi.org/10.1038/s41586-022-05442-1}{Nature {\bf
  613}, 468--473}~(2023).

\bibitem{ho2022exact}
Wen~Wei Ho and Soonwon Choi.
\newblock ``Exact emergent quantum state designs from quantum chaotic
  dynamics''.
\newblock \href{https://dx.doi.org/10.1103/PhysRevLett.128.060601}{Phys. Rev.
  Lett. {\bf 128}, 060601}~(2022).

\bibitem{claeys2022emergent}
Pieter~W Claeys and Austen Lamacraft.
\newblock ``Emergent quantum state designs and biunitarity in dual-unitary
  circuit dynamics''.
\newblock
  \href{https://dx.doi.org/https://doi.org/10.22331/q-2022-06-15-738}{Quantum
  {\bf 6}, 738}~(2022).

\bibitem{ippoliti2022solvable}
Matteo Ippoliti and Wen~Wei Ho.
\newblock ``Solvable model of deep thermalization with distinct design times''.
\newblock \href{https://dx.doi.org/10.22331/q-2022-12-29-886}{Quantum {\bf 6},
  886}~(2022).

\bibitem{lucas2022generalized}
Maxime Lucas, Lorenzo Piroli, Jacopo De~Nardis, and Andrea De~Luca.
\newblock ``Generalized deep thermalization for free fermions''.
\newblock \href{https://dx.doi.org/10.1103/PhysRevA.107.032215}{Phys. Rev. A
  {\bf 107}, 032215}~(2023).

\bibitem{bhore2023deep}
Tanmay Bhore, Jean-Yves Desaules, and Zlatko Papi\ifmmode~\acute{c}\else
  \'{c}\fi{}.
\newblock ``Deep thermalization in constrained quantum systems''.
\newblock \href{https://dx.doi.org/10.1103/PhysRevB.108.104317}{Phys. Rev. B
  {\bf 108}, 104317}~(2023).

\bibitem{mcginley2022shadow}
Max McGinley and Michele Fava.
\newblock ``Shadow tomography from emergent state designs in analog quantum
  simulators''.
\newblock \href{https://dx.doi.org/10.1103/PhysRevLett.131.160601}{Phys. Rev.
  Lett. {\bf 131}, 160601}~(2023).

\bibitem{pilatowsky2023complete}
Sa\'ul Pilatowsky-Cameo, Ceren~B. Dag, Wen~Wei Ho, and Soonwon Choi.
\newblock ``Complete hilbert-space ergodicity in quantum dynamics of
  generalized fibonacci drives''.
\newblock \href{https://dx.doi.org/10.1103/PhysRevLett.131.250401}{Phys. Rev.
  Lett. {\bf 131}, 250401}~(2023).

\bibitem{mark2024maximum}
Daniel~K Mark, Federica Surace, Andreas Elben, Adam~L Shaw, Joonhee Choi, Gil
  Refael, Manuel Endres, and Soonwon Choi.
\newblock ``A maximum entropy principle in deep thermalization and in
  hilbert-space ergodicity''~(2024).
\newblock  \href{http://arxiv.org/abs/2403.11970}{arXiv:2403.11970}.

\bibitem{dymarsky2016subsystem}
Anatoly Dymarsky, Nima Lashkari, and Hong Liu.
\newblock ``Subsystem eigenstate thermalization hypothesis''.
\newblock \href{https://dx.doi.org/10.1103/PhysRevE.97.012140}{Phys. Rev. E
  {\bf 97}, 012140}~(2018).

\bibitem{huang2021universal}
Yichen Huang.
\newblock ``Universal entanglement of mid-spectrum eigenstates of chaotic local
  hamiltonians''.
\newblock \href{https://dx.doi.org/10.1016/j.nuclphysb.2021.115373}{Nuclear
  Physics B {\bf 966}, 115373}~(2021).

\bibitem{lu2019renyi}
Tsung-Cheng Lu and Tarun Grover.
\newblock ``Renyi entropy of chaotic eigenstates''.
\newblock \href{https://dx.doi.org/10.1103/PhysRevE.99.032111}{Phys. Rev. E
  {\bf 99}, 032111}~(2019).

\bibitem{murthy2019structure}
Chaitanya Murthy and Mark Srednicki.
\newblock ``Structure of chaotic eigenstates and their entanglement entropy''.
\newblock \href{https://dx.doi.org/10.1103/PhysRevE.100.022131}{Physical Review
  E {\bf 100}, 022131}~(2019).

\bibitem{shi2023local}
Zhengyan~Darius Shi, Shreya Vardhan, and Hong Liu.
\newblock ``Local dynamics and the structure of chaotic eigenstates''.
\newblock \href{https://dx.doi.org/10.1103/PhysRevB.108.224305}{Physical Review
  B {\bf 108}, 224305}~(2023).

\bibitem{hahn2023statistical}
Dominik Hahn, David~J. Luitz, and J.~T. Chalker.
\newblock ``Eigenstate correlations, the eigenstate thermalization hypothesis,
  and quantum information dynamics in chaotic many-body quantum systems''.
\newblock \href{https://dx.doi.org/10.1103/PhysRevX.14.031029}{Phys. Rev. X
  {\bf 14}, 031029}~(2024).

\bibitem{jindal2024generalized}
Siddharth Jindal and Pavan Hosur.
\newblock ``Generalized free cumulants for quantum chaotic systems''~(2024).
\newblock  \href{http://arxiv.org/abs/2401.13829}{arXiv:2401.13829}.

\bibitem{doyon2020fluctuations}
Benjamin Doyon and Jason Myers.
\newblock ``Fluctuations in ballistic transport from euler hydrodynamics''.
\newblock In Annales Henri Poincar{\'e}.
\newblock \href{https://dx.doi.org/10.1007/s00023-019-00860-w}{Volume~21, pages
  255--302}.
\newblock Springer~(2020).

\bibitem{myers2020transport}
Jason Myers, Joe Bhaseen, Rosemary~J Harris, and Benjamin Doyon.
\newblock ``Transport fluctuations in integrable models out of equilibrium''.
\newblock \href{https://dx.doi.org/10.21468/SciPostPhys.8.1.007}{SciPost
  Physics {\bf 8}, 007}~(2020).

\bibitem{fava2021hydrodynamic}
Michele Fava, Sounak Biswas, Sarang Gopalakrishnan, Romain Vasseur, and
  SA~Parameswaran.
\newblock ``Hydrodynamic nonlinear response of interacting integrable
  systems''.
\newblock \href{https://dx.doi.org/10.1073/pnas.2106945118}{Proceedings of the
  National Academy of Sciences {\bf 118}, e2106945118}~(2021).

\bibitem{delacretaz2024nonlinear}
Luca~V Delacr{\'e}taz and Ruchira Mishra.
\newblock ``Nonlinear response in diffusive systems''.
\newblock \href{https://dx.doi.org/10.21468/SciPostPhys.16.2.047}{SciPost
  Physics {\bf 16}, 047}~(2024).

\bibitem{kawamoto2024strategy}
Taishi Kawamoto.
\newblock ``A strategy for proving the strong eigenstate thermalization
  hypothesis: Chaotic systems and holography''~(2024).
\newblock  \href{http://arxiv.org/abs/2411.09746}{arXiv:2411.09746}.

\bibitem{maldacena2016bound}
Juan Maldacena, Stephen~H Shenker, and Douglas Stanford.
\newblock ``A bound on chaos''.
\newblock \href{https://dx.doi.org/10.1007/JHEP08(2016)106}{Journal of High
  Energy Physics {\bf 2016}, 106}~(2016).

\bibitem{hosur2016chaos}
Pavan Hosur, Xiao-Liang Qi, Daniel~A Roberts, and Beni Yoshida.
\newblock ``Chaos in quantum channels''.
\newblock \href{https://dx.doi.org/10.1007/JHEP02(2016)004}{Journal of High
  Energy Physics {\bf 2016}, 004}~(2016).

\bibitem{xu2022scrambling}
Shenglong Xu and Brian Swingle.
\newblock ``Scrambling dynamics and out-of-time ordered correlators in quantum
  many-body systems: a tutorial''~(2022).
\newblock  \href{http://arxiv.org/abs/2202.07060}{arXiv:2202.07060}.

\bibitem{garcia2022out}
Ignacio García-Mata, Rodolfo~A. Jalabert, and Diego~A. Wisniacki.
\newblock ``Out-of-time-order correlators and quantum chaos''~(2022).
\newblock  \href{http://arxiv.org/abs/2209.07965}{arXiv:2209.07965}.

\bibitem{larkin1969quasiclassical}
A.~I. Larkin and Yu.~N. Ovchinnikov.
\newblock ``Quasiclassical method in the theory of superconductivity''.
\newblock Soviet Physics JETP {\bf 28}, 1200--1205~(1969).
\newblock  url:~\url{https://jetp.ras.ru/cgi-bin/dn/e_028_06_1200.pdf}.

\bibitem{KitaTalk}
A.~Kitaev.
\newblock ``Talk given at the fundamental physics prize symposium''.
\newblock YouTube. ~(20145).
\newblock  url:~\url{https://www.youtube.com/watch?v=OQ9qN8j7EZI}.

\bibitem{roberts2017chaos}
Daniel~A Roberts and Beni Yoshida.
\newblock ``Chaos and complexity by design''.
\newblock \href{https://dx.doi.org/10.1007/JHEP04(2017)121}{Journal of High
  Energy Physics {\bf 2017}, 121}~(2017).

\bibitem{cotler2017chaos}
Jordan Cotler, Nicholas Hunter-Jones, Junyu Liu, and Beni Yoshida.
\newblock ``Chaos, complexity, and random matrices''.
\newblock \href{https://dx.doi.org/10.1007/JHEP11(2017)048}{Journal of High
  Energy Physics {\bf 2017}, 1--60}~(2017).

\bibitem{tsuji2018bound}
Naoto Tsuji, Tomohiro Shitara, and Masahito Ueda.
\newblock ``Bound on the exponential growth rate of out-of-time-ordered
  correlators''.
\newblock \href{https://dx.doi.org/10.1103/PhysRevE.98.012216}{Physical Review
  E {\bf 98}, 012216}~(2018).

\bibitem{cotler2020spectral}
Jordan Cotler and Nicholas Hunter-Jones.
\newblock ``Spectral decoupling in many-body quantum chaos''.
\newblock \href{https://dx.doi.org/10.1007/JHEP12(2020)205}{Journal of High
  Energy Physics {\bf 2020}, 1--62}~(2020).

\bibitem{leone2021quantum}
Lorenzo Leone, Salvatore~FE Oliviero, You Zhou, and Alioscia Hamma.
\newblock ``Quantum chaos is quantum''.
\newblock \href{https://dx.doi.org/10.22331/q-2021-05-04-453}{Quantum {\bf 5},
  453}~(2021).

\bibitem{Pappalardi2023quantum}
Silvia Pappalardi and Jorge Kurchan.
\newblock ``Quantum bounds on the generalized lyapunov exponents''.
\newblock \href{https://dx.doi.org/10.3390/e25020246}{Entropy {\bf 25},
  246}~(2023).

\bibitem{trunin2023quantum}
Dmitrii~A Trunin.
\newblock ``Quantum chaos without false positives''.
\newblock \href{https://dx.doi.org/10.1103/PhysRevD.108.L101703}{Physical
  Review D {\bf 108}, L101703}~(2023).

\bibitem{trunin2023refined}
Dmitrii~A Trunin.
\newblock ``Refined quantum lyapunov exponents from replica out-of-time-order
  correlators''.
\newblock \href{https://dx.doi.org/10.1103/PhysRevD.108.105023}{Physical Review
  D {\bf 108}, 105023}~(2023).

\bibitem{srednicki1994chaos}
Mark Srednicki.
\newblock ``Chaos and quantum thermalization''.
\newblock \href{https://dx.doi.org/10.1103/PhysRevE.50.888}{Physical Review E
  {\bf 50}, 888--901}~(1994).

\bibitem{srednicki1999approach}
Mark Srednicki.
\newblock ``The approach to thermal equilibrium in quantized chaotic systems''.
\newblock \href{https://dx.doi.org/10.1088/0305-4470/32/7/007}{Journal of
  Physics A: Mathematical and General {\bf 32}, 1163--1175}~(1999).

\bibitem{dalessio2016from}
Luca D'Alessio, Yariv Kafri, Anatoli Polkovnikov, and Marcos Rigol.
\newblock ``From quantum chaos and eigenstate thermalization to statistical
  mechanics and thermodynamics''.
\newblock \href{https://dx.doi.org/10.1080/00018732.2016.1198134}{Advances in
  Physics {\bf 65}, 239--362}~(2016).

\bibitem{foini2019eigenstate}
Laura Foini and Jorge Kurchan.
\newblock ``Eigenstate thermalization hypothesis and out of time order
  correlators''.
\newblock \href{https://dx.doi.org/10.1103/PhysRevE.99.042139}{Physical Review
  E {\bf 99}, 042139}~(2019).

\bibitem{prosen1999}
Tomaz Prosen.
\newblock ``Ergodic properties of a generic nonintegrable quantum many-body
  system in the thermodynamic limit''.
\newblock \href{https://dx.doi.org/10.1103/PhysRevE.60.3949}{Phys. Rev. E {\bf
  60}, 3949--3968}~(1999).

\bibitem{sonner2017eigenstate}
Julian Sonner and Manuel Vielma.
\newblock ``Eigenstate thermalization in the sachdev-ye- model''.
\newblock \href{https://dx.doi.org/10.1007/JHEP11(2017)149}{Journal of High
  Energy Physics {\bf 2017}, 1--28}~(2017).

\bibitem{foini2019eigenstate2}
Laura Foini and Jorge Kurchan.
\newblock ``Eigenstate thermalization and rotational invariance in ergodic
  quantum systems''.
\newblock \href{https://dx.doi.org/10.1103/PhysRevLett.123.260601}{Phys. Rev.
  Lett. {\bf 123}, 260601}~(2019).

\bibitem{chan2019eigenstate}
Amos Chan, Andrea De~Luca, and J.~T. Chalker.
\newblock ``Eigenstate correlations, thermalization, and the butterfly
  effect''.
\newblock \href{https://dx.doi.org/10.1103/PhysRevLett.122.220601}{Phys. Rev.
  Lett. {\bf 122}, 220601}~(2019).

\bibitem{murthy2019bounds}
Chaitanya Murthy and Mark Srednicki.
\newblock ``Bounds on chaos from the eigenstate thermalization hypothesis''.
\newblock \href{https://dx.doi.org/10.1103/PhysRevLett.123.230606}{Physical
  Review Letters {\bf 123}, 230606}~(2019).

\bibitem{richter2020eigenstate}
Jonas Richter, Anatoly Dymarsky, Robin Steinigeweg, and Jochen Gemmer.
\newblock ``Eigenstate thermalization hypothesis beyond standard indicators:
  Emergence of random-matrix behavior at small frequencies''.
\newblock \href{https://dx.doi.org/10.1103/PhysRevE.102.042127}{Physical Review
  E {\bf 102}, 042127}~(2020).

\bibitem{wang2021eigenstate}
Jiaozi Wang, Mats~H Lamann, Jonas Richter, Robin Steinigeweg, Anatoly Dymarsky,
  and Jochen Gemmer.
\newblock ``Eigenstate thermalization hypothesis and its deviations from
  random-matrix theory beyond the thermalization time''~(2021).
\newblock  \href{http://arxiv.org/abs/2110.04085}{arXiv:2110.04085}.

\bibitem{brenes2021out}
Marlon Brenes, Silvia Pappalardi, Mark~T Mitchison, John Goold, and Alessandro
  Silva.
\newblock ``Out-of-time-order correlations and the fine structure of eigenstate
  thermalization''.
\newblock \href{https://dx.doi.org/10.1103/PhysRevE.104.034120}{Physical Review
  E {\bf 104}, 034120}~(2021).

\bibitem{dymarsky2022bound}
Anatoly Dymarsky.
\newblock ``Bound on eigenstate thermalization from transport''.
\newblock \href{https://dx.doi.org/10.1103/PhysRevLett.128.190601}{Phys. Rev.
  Lett. {\bf 128}, 190601}~(2022).

\bibitem{nussinov2022exact}
Zohar Nussinov and Saurish Chakrabarty.
\newblock ``Exact universal chaos, speed limit, acceleration, planckian
  transport coefficient,“collapse” to equilibrium, and other bounds in
  thermal quantum systems''.
\newblock \href{https://dx.doi.org/10.1016/j.aop.2022.168970}{Annals of Physics
  {\bf 443}, 168970}~(2022).

\bibitem{jafferis2022matrix}
Daniel~Louis Jafferis, David~K Kolchmeyer, Baur Mukhametzhanov, and Julian
  Sonner.
\newblock ``Matrix models for eigenstate thermalization''~(2022).
\newblock  \href{http://arxiv.org/abs/2209.02130}{arXiv:2209.02130}.

\bibitem{jafferis2022jt}
Daniel~Louis Jafferis, David~K Kolchmeyer, Baur Mukhametzhanov, and Julian
  Sonner.
\newblock ``Jt gravity with matter, generalized eth, and random
  matrices''~(2022).
\newblock  \href{http://arxiv.org/abs/2209.02131}{arXiv:2209.02131}.

\bibitem{wang2023emergence}
Jiaozi Wang, Jonas Richter, Mats~H Lamann, Robin Steinigeweg, Jochen Gemmer,
  and Anatoly Dymarsky.
\newblock ``Emergence of unitary symmetry of microcanonically truncated
  operators in chaotic quantum systems''.
\newblock \href{https://dx.doi.org/10.1103/PhysRevE.110.L032203}{Physical
  Review E {\bf 110}, L032203}~(2024).

\bibitem{fava2023designs}
Michele Fava, Jorge Kurchan, and Silvia Pappalardi.
\newblock ``Designs via free probability''.
\newblock \href{https://dx.doi.org/10.1103/PhysRevX.15.011031}{Physical Review
  X {\bf 15}, 011031}~(2025).

\bibitem{pappalardi2023general}
Silvia Pappalardi, Felix Fritzsch, and Toma{\v{z}} Prosen.
\newblock ``Full eigenstate thermalization via free cumulants in quantum
  lattice systems''~(2023).
\newblock  \href{http://arxiv.org/abs/2303.00713}{arXiv:2303.00713}.

\bibitem{pappalardi2024microcanonical}
Silvia Pappalardi, Laura Foini, and Jorge Kurchan.
\newblock ``Microcanonical windows on quantum operators''.
\newblock \href{https://dx.doi.org/10.22331/q-2024-01-11-1227}{Quantum {\bf 8},
  1227}~(2024).

\bibitem{fritzsch2024microcanonical}
Felix Fritzsch, Toma{\v{z}} Prosen, and Silvia Pappalardi.
\newblock ``Microcanonical free cumulants in lattice systems''.
\newblock \href{https://dx.doi.org/10.1103/PhysRevB.111.054303}{Physical Review
  B {\bf 111}, 054303}~(2025).

\bibitem{pappalardi2022eigenstate}
Silvia Pappalardi, Laura Foini, and Jorge Kurchan.
\newblock ``Eigenstate thermalization hypothesis and free probability''.
\newblock \href{https://dx.doi.org/10.1103/PhysRevLett.129.170603}{Physical
  Review Letters {\bf 129}, 170603}~(2022).

\bibitem{voiculescu1991limit}
Dan Voiculescu.
\newblock ``Limit laws for random matrices and free products''.
\newblock \href{https://dx.doi.org/10.1007/BF01245072}{Inventiones Mathematicae
  {\bf 104}, 201--220}~(1991).

\bibitem{voiculescu1992free}
Dan~V Voiculescu, Ken~J Dykema, and Alexandru Nica.
\newblock ``Free random variables''.
\newblock \href{https://dx.doi.org/10.1090/crmm/001}{American Mathematical
  Society}. ~(1992).

\bibitem{speicher1997free}
Roland Speicher.
\newblock ``Free probability theory and non-crossing partitions''.
\newblock S{\'e}minaire Lotharingien de Combinatoire{\bf 39}~(1997).
\newblock  url:~\url{https://www.emis.de/journals/SLC/wpapers/s39speicher.pdf}.

\bibitem{brezin1978planar}
Edouard Br{\'e}zin, Claude Itzykson, Giorgio Parisi, and Jean-Bernard Zuber.
\newblock ``Planar diagrams''.
\newblock \href{https://dx.doi.org/10.1007/BF01614153}{Communications in
  Mathematical Physics {\bf 59}, 35--51}~(1978).

\bibitem{cvitanovic1981planar}
Predrag Cvitanovi{\'c}.
\newblock ``Planar perturbation expansion''.
\newblock \href{https://dx.doi.org/10.1016/0370-2693(81)90801-7}{Physics
  Letters B {\bf 99}, 49--52}~(1981).

\bibitem{cvitanovic1982planar}
Predrag Cvitanovi{\'c}, P~G Lauwers, and P~N Scharbach.
\newblock ``The planar sector of field theories''.
\newblock \href{https://dx.doi.org/10.1016/0550-3213(82)90320-0}{Nuclear
  Physics B {\bf 203}, 385--412}~(1982).

\bibitem{collins2016random}
Beno{\^\i}t Collins and Ion Nechita.
\newblock ``Random matrix techniques in quantum information theory''.
\newblock \href{https://dx.doi.org/10.1063/1.4936880}{Journal of Mathematical
  Physics {\bf 57}, 015215}~(2016).

\bibitem{collins2010random}
Beno{\^\i}t Collins and Ion Nechita.
\newblock ``Random quantum channels i: Graphical calculus and the bell state
  phenomenon''.
\newblock \href{https://dx.doi.org/10.1007/s00220-010-1012-0}{Communications in
  Mathematical Physics {\bf 297}, 345--370}~(2010).

\bibitem{kutlerflam2021distinguishing}
Jonah Kudler-Flam, Vladimir Narovlansky, and Shinsei Ryu.
\newblock ``Distinguishing random and black hole microstates''.
\newblock \href{https://dx.doi.org/10.1103/PRXQuantum.2.040340}{PRX Quantum
  {\bf 2}, 040340}~(2021).

\bibitem{kudler2022negativity}
Jonah Kudler-Flam, Vladimir Narovlansky, and Shinsei Ryu.
\newblock ``Negativity spectra in random tensor networks and holography''.
\newblock \href{https://dx.doi.org/10.1007/JHEP02(2022)076}{Journal of High
  Energy Physics {\bf 2022}, 1--74}~(2022).

\bibitem{cheng2024random}
Newton Cheng, C{\'e}cilia Lancien, Geoff Penington, Michael Walter, and Freek
  Witteveen.
\newblock ``Random tensor networks with non-trivial links''.
\newblock In Annales Henri Poincar{\'e}.
\newblock \href{https://dx.doi.org/10.1007/s00023-023-01358-2}{Volume~25, pages
  2107--2212}.
\newblock Springer~(2024).

\bibitem{movassagh2010isotropic}
Ramis Movassagh and Alan Edelman.
\newblock ``Isotropic entanglement''~(2010).
\newblock  \href{http://arxiv.org/abs/1012.5039}{arXiv:1012.5039}.

\bibitem{movassagh2011density}
Ramis Movassagh and Alan Edelman.
\newblock ``Density of states of quantum spin systems from isotropic
  entanglement''.
\newblock \href{https://dx.doi.org/10.1103/PhysRevLett.107.097205}{Physical
  Review Letters {\bf 107}, 097205}~(2011).

\bibitem{chen2012error}
Jiahao Chen, Eric Hontz, Jeremy Moix, Matthew Welborn, Troy Van~Voorhis,
  Alberto Su\'arez, Ramis Movassagh, and Alan Edelman.
\newblock ``Error analysis of free probability approximations to the density of
  states of disordered systems''.
\newblock \href{https://dx.doi.org/10.1103/PhysRevLett.109.036403}{Phys. Rev.
  Lett. {\bf 109}, 036403}~(2012).

\bibitem{hruza2023coherent}
Ludwig Hruza and Denis Bernard.
\newblock ``Coherent fluctuations in noisy mesoscopic systems, the open quantum
  {SSEP}, and free probability''.
\newblock \href{https://dx.doi.org/10.1103/physrevx.13.011045}{Physical Review
  X{\bf 13}}~(2023).

\bibitem{bauer2023bernoulli}
Michel Bauer, Denis Bernard, Philippe Biane, and Ludwig Hruza.
\newblock ``Bernoulli variables, classical exclusion processes and free
  probability''.
\newblock \href{https://dx.doi.org/10.1007/s00023-023-01320-2}{Annales Henri
  Poincaré {\bf 24}, 1--48}~(2023).

\bibitem{bernard2023exact}
Denis Bernard and Ludwig Hruza.
\newblock ``Exact entanglement in the driven quantum symmetric simple exclusion
  process''.
\newblock \href{https://dx.doi.org/10.21468/SciPostPhys.15.4.175}{SciPost
  Physics {\bf 15}, 175}~(2023).

\bibitem{berkooz2019towards}
Micha Berkooz, Mikhail Isachenkov, Vladimir Narovlansky, and Genis Torrents.
\newblock ``Towards a full solution of the large n double-scaled syk model''.
\newblock \href{https://dx.doi.org/10.1007/JHEP03(2019)079}{Journal of High
  Energy Physics {\bf 2019}, 079}~(2019).

\bibitem{penington2022replica}
Geoff Penington, Stephen~H Shenker, Douglas Stanford, and Zhenbin Yang.
\newblock ``Replica wormholes and the black hole interior''.
\newblock \href{https://dx.doi.org/10.1007/JHEP03(2022)205}{Journal of High
  Energy Physics {\bf 2022}, 205}~(2022).

\bibitem{wang2023beyond}
Jinzhao Wang.
\newblock ``Beyond islands: a free probabilistic approach''.
\newblock \href{https://dx.doi.org/10.1007/JHEP10(2023)040}{Journal of High
  Energy Physics {\bf 2023}, 40}~(2023).

\bibitem{wu2024non}
Shuang Wu.
\newblock ``Non-commutative probability insights into the double-scaling limit
  syk model with constant perturbations: moments, cumulants and
  q-independence''.
\newblock
  \href{https://dx.doi.org/https://doi.org/10.1088/1751-8121/ad65a6}{Journal of
  Physics A: Mathematical and Theoretical {\bf 57}, 325203}~(2024).

\bibitem{chandrasekaran2023large}
Venkatesa Chandrasekaran, Geoff Penington, and Edward Witten.
\newblock ``Large n algebras and generalized entropy''.
\newblock \href{https://dx.doi.org/10.1007/JHEP04(2023)009}{Journal of High
  Energy Physics {\bf 2023}, 009}~(2023).

\bibitem{cipolloni2022thermalisation}
Giorgio Cipolloni, L{\'a}szl{\'o} Erd{\H{o}}s, and Dominik Schr{\"o}der.
\newblock ``Thermalisation for wigner matrices''.
\newblock
  \href{https://dx.doi.org/https://doi.org/10.1016/j.jfa.2022.109394}{Journal
  of Functional Analysis {\bf 282}, 109394}~(2022).

\bibitem{chen2024free}
Hyaline~Junhe Chen and Jonah Kudler-Flam.
\newblock ``Free independence and the noncrossing partition lattice in
  dual-unitary quantum circuits''~(2024).
\newblock  \href{http://arxiv.org/abs/2409.17226}{arXiv:2409.17226}.

\bibitem{haake1987classical}
F.~Haake, M.~Kuś, and R.~Scharf.
\newblock ``Classical and quantum chaos for a kicked top''.
\newblock \href{https://dx.doi.org/10.1007/BF01303727}{Z. Phys. B-Condensed
  Matter {\bf 65}, 381--395}~(1987).

\bibitem{kus1987symmetry}
M.~Kuś, R.~Scharf, and F.~Haake.
\newblock ``Symmetry versus degree of level repulsion for kicked quantum
  systems''.
\newblock \href{https://dx.doi.org/10.1007/BF01312770}{Zeitschrift für Physik
  B Condensed Matter {\bf 66}, 129--134}~(1987).

\bibitem{benzi1984multifractal}
Roberto Benzi, Giovanni Paladin, Giorgio Parisi, and Angelo Vulpiani.
\newblock ``On the multifractal nature of fully developed turbulence and
  chaotic systems''.
\newblock
  \href{https://dx.doi.org/https://doi.org/10.1142/9789812799050_0017}{Journal
  of Physics A: Mathematical and General {\bf 17}, 3521}~(1984).

\bibitem{paladin1987anomalous}
Giovanni Paladin and Angelo Vulpiani.
\newblock ``Anomalous scaling laws in multifractal objects''.
\newblock \href{https://dx.doi.org/10.1016/0370-1573(87)90110-4}{Physics
  Reports {\bf 156}, 147--225}~(1987).

\bibitem{paladin1986intermittency}
Giovanni Paladin and Angelo Vulpiani.
\newblock ``Intermittency in chaotic systems and r{\'e}nyi entropies''.
\newblock \href{https://dx.doi.org/10.1088/0305-4470/19/16/009}{Journal of
  Physics A: Mathematical and General {\bf 19}, L997}~(1986).

\bibitem{crisanti1988lyapunov}
Andrea Crisanti, Giovanni Paladin, and Angelo Vulpiani.
\newblock ``Generalized lyapunov exponents in high-dimensional chaotic dynamics
  and products of large random matrices''.
\newblock \href{https://dx.doi.org/10.1007/BF01014215}{Journal of Statistical
  Physics {\bf 53}, 583--601}~(1988).

\bibitem{evers200fluctuations}
F.~Evers and A.~D. Mirlin.
\newblock ``Fluctuations of the inverse participation ratio at the anderson
  transition''.
\newblock \href{https://dx.doi.org/10.1103/PhysRevLett.84.3690}{Phys. Rev.
  Lett. {\bf 84}, 3690--3693}~(2000).

\bibitem{mace2019multifractal}
Nicolas Mac\'e, Fabien Alet, and Nicolas Laflorencie.
\newblock ``Multifractal scalings across the many-body localization
  transition''.
\newblock \href{https://dx.doi.org/10.1103/PhysRevLett.123.180601}{Phys. Rev.
  Lett. {\bf 123}, 180601}~(2019).

\bibitem{sierant2022universal}
Piotr Sierant and Xhek Turkeshi.
\newblock ``Universal behavior beyond multifractality of wave functions at
  measurement-induced phase transitions''.
\newblock \href{https://dx.doi.org/10.1103/PhysRevLett.128.130605}{Phys. Rev.
  Lett. {\bf 128}, 130605}~(2022).

\bibitem{zenodo2024}
Elisa Vallini and Silvia Pappalardi.
\newblock ``Codes and data for: {Long-time Freeness in the Kicked Top}''.
\newblock \href{https://dx.doi.org/10.5281/zenodo.19698543}{Zenodo}~(2026).

\bibitem{chaudhury2009quantum}
S.~Chaudhury, A.~Smith, B.~E. Anderson, S.~Ghose, and P.~S. Jessen.
\newblock ``Quantum signatures of chaos in a kicked top''.
\newblock \href{https://dx.doi.org/10.1038/nature08396}{Nature {\bf 461},
  768--771}~(2009).

\bibitem{wang2021multifractality}
Qian Wang and Marko Robnik.
\newblock ``Multifractality in quasienergy space of coherent states as a
  signature of quantum chaos''.
\newblock \href{https://dx.doi.org/10.3390/e23101347}{Entropy {\bf 23},
  1347}~(2021).

\bibitem{pappalardi2018scrambling}
Silvia Pappalardi, Angelo Russomanno, Bojan \ifmmode \check{Z}\else
  \v{Z}\fi{}unkovi\ifmmode~\check{c}\else \v{c}\fi{}, Fernando Iemini,
  Alessandro Silva, and Rosario Fazio.
\newblock ``Scrambling and entanglement spreading in long-range spin chains''.
\newblock \href{https://dx.doi.org/10.1103/PhysRevB.98.134303}{Phys. Rev. B
  {\bf 98}, 134303}~(2018).

\bibitem{seshadri2018tripartite}
Akshay Seshadri, Vaibhav Madhok, and Arul Lakshminarayan.
\newblock ``Tripartite mutual information, entanglement, and scrambling in
  permutation symmetric systems with an application to quantum chaos''.
\newblock \href{https://dx.doi.org/10.1103/PhysRevE.98.052205}{Phys. Rev. E
  {\bf 98}, 052205}~(2018).

\bibitem{sieberer2019digital}
Lukas~M. Sieberer, Tobias Olsacher, Andreas Elben, Markus Heyl, Philipp Hauke,
  Fritz Haake, and Peter Zoller.
\newblock ``Digital quantum simulation, trotter errors, and quantum chaos of
  the kicked top''.
\newblock \href{https://dx.doi.org/10.1038/s41534-019-0192-5}{npj Quantum
  Information{\bf 5}}~(2019).

\bibitem{pilatowsky2020positive}
Sa\'ul Pilatowsky-Cameo, Jorge Ch\'avez-Carlos, Miguel~A. Bastarrachea-Magnani,
  Pavel Str\'ansk\'y, Sergio Lerma-Hern\'andez, Lea~F. Santos, and Jorge~G.
  Hirsch.
\newblock ``Positive quantum lyapunov exponents in experimental systems with a
  regular classical limit''.
\newblock \href{https://dx.doi.org/10.1103/PhysRevE.101.010202}{Phys. Rev. E
  {\bf 101}, 010202}~(2020).

\bibitem{lerose2020bridging}
Alessio Lerose and Silvia Pappalardi.
\newblock ``Bridging entanglement dynamics and chaos in semiclassical
  systems''.
\newblock \href{https://dx.doi.org/10.1103/PhysRevA.102.032404}{Phys. Rev. A
  {\bf 102}, 032404}~(2020).

\bibitem{deutsch1991quantum}
J.~M. Deutsch.
\newblock ``Quantum statistical mechanics in a closed system''.
\newblock \href{https://dx.doi.org/10.1103/PhysRevA.43.2046}{Physical Review A
  {\bf 43}, 2046--2049}~(1991).

\bibitem{speicher2016free}
Roland Speicher.
\newblock ``Free probability theory''.
\newblock \href{https://dx.doi.org/10.1365/s13291-016-0150-5}{Jahresbericht der
  Deutschen Mathematiker-Vereinigung{\bf 119}}~(2016).

\bibitem{xia2019simple}
Xiang-Gen Xia.
\newblock ``A simple introduction to free probability theory and its
  application to random matrices''~(2019).
\newblock  \href{http://arxiv.org/abs/1902.10763}{arXiv:1902.10763}.

\bibitem{mingo2017free}
James~A Mingo and Roland Speicher.
\newblock ``Free probability and random matrices''.
\newblock \href{https://dx.doi.org/10.1007/978-1-4939-6942-5}{Springer}.
  ~(2017).

\bibitem{dowling2025free}
Neil Dowling, Jacopo~De Nardis, Markus Heinrich, Xhek Turkeshi, and Silvia
  Pappalardi.
\newblock ``Free independence and unitary design from random matrix product
  unitaries''~(2025).
\newblock  \href{http://arxiv.org/abs/2508.00051}{arXiv:2508.00051}.

\bibitem{linden2009quantum}
Noah Linden, Sandu Popescu, Anthony~J Short, and Andreas Winter.
\newblock ``Quantum mechanical evolution towards thermal equilibrium''.
\newblock \href{https://dx.doi.org/10.1103/PhysRevE.79.061103}{Physical Review
  E—Statistical, Nonlinear, and Soft Matter Physics {\bf 79}, 061103}~(2009).

\bibitem{speicher2025lecturey}
Roland Speicher.
\newblock ``Lecture notes on "free probability theory"''~(2025).
\newblock  \href{http://arxiv.org/abs/1908.08125}{arXiv:1908.08125}.

\bibitem{huang2019finite}
Yichen Huang, Fernando G. S.~L. Brand\~ao, and Yong-Liang Zhang.
\newblock ``Finite-size scaling of out-of-time-ordered correlators at late
  times''.
\newblock \href{https://dx.doi.org/10.1103/PhysRevLett.123.010601}{Phys. Rev.
  Lett. {\bf 123}, 010601}~(2019).

\bibitem{fritzsch2021eigenstate}
Felix Fritzsch and Tomaz Prosen.
\newblock ``Eigenstate thermalization in dual-unitary quantum circuits:
  Asymptotics of spectral functions''.
\newblock \href{https://dx.doi.org/10.1103/PhysRevE.103.062133}{Phys. Rev. E
  {\bf 103}, 062133}~(2021).

\bibitem{garcia2018chaos}
Ignacio Garc\'{\i}a-Mata, Marcos Saraceno, Rodolfo~A. Jalabert, Augusto~J.
  Roncaglia, and Diego~A. Wisniacki.
\newblock ``Chaos signatures in the short and long time behavior of the
  out-of-time ordered correlator''.
\newblock \href{https://dx.doi.org/10.1103/PhysRevLett.121.210601}{Phys. Rev.
  Lett. {\bf 121}, 210601}~(2018).

\bibitem{notenson2023classical}
Tom{\'a}s Notenson, Ignacio Garc{\'\i}a-Mata, Augusto~J Roncaglia, and Diego~A
  Wisniacki.
\newblock ``Classical approach to equilibrium of out-of-time ordered
  correlators in mixed systems''.
\newblock \href{https://dx.doi.org/10.1103/PhysRevE.107.064207}{Physical Review
  E {\bf 107}, 064207}~(2023).

\bibitem{kobrin2021many}
Bryce Kobrin, Zhenbin Yang, Gregory~D. Kahanamoku-Meyer, Christopher~T. Olund,
  Joel~E. Moore, Douglas Stanford, and Norman~Y. Yao.
\newblock ``Many-body chaos in the sachdev-ye-kitaev model''.
\newblock \href{https://dx.doi.org/10.1103/PhysRevLett.126.030602}{Phys. Rev.
  Lett. {\bf 126}, 030602}~(2021).

\bibitem{polchinski2015chaos}
Joseph Polchinski.
\newblock ``Chaos in the black hole s-matrix''~(2015).
\newblock  \href{http://arxiv.org/abs/1505.08108}{arXiv:1505.08108}.

\bibitem{fritzsch2025free}
Felix Fritzsch and Pieter~W. Claeys.
\newblock ``Free probability in a minimal quantum circuit model''~(2025).
\newblock  \href{http://arxiv.org/abs/2506.11197}{arXiv:2506.11197}.

\bibitem{vardhan2025free}
Shreya Vardhan and Jinzhao Wang.
\newblock ``Free mutual information and higher-point otocs''~(2025).
\newblock  \href{http://arxiv.org/abs/2509.13406}{arXiv:2509.13406}.

\bibitem{fritzsch2025free2}
Felix Fritzsch, Gabriel~O. Alves, Michael~A. Rampp, and Pieter~W. Claeys.
\newblock ``Free cumulants and full eigenstate thermalization from boundary
  scrambling''~(2025).
\newblock  \href{http://arxiv.org/abs/2509.08060}{arXiv:2509.08060}.

\bibitem{jahnke2025free}
Viktor Jahnke, Pratik Nandy, Kuntal Pal, Hugo~A. Camargo, and Keun-Young Kim.
\newblock ``Free probability approach to spectral and operator statistics in
  rosenzweig-porter random matrix ensembles''.
\newblock \href{https://dx.doi.org/10.1007/jhep12(2025)002}{Journal of High
  Energy Physics{\bf 2025}}~(2025).

\bibitem{cotler2017black}
Jordan~S Cotler, Guy Gur-Ari, Masanori Hanada, Joseph Polchinski, Phil Saad,
  Stephen~H Shenker, Douglas Stanford, Alexandre Streicher, and Masaki Tezuka.
\newblock ``Black holes and random matrices''.
\newblock \href{https://dx.doi.org/10.1007/JHEP05(2017)118}{Journal of High
  Energy Physics {\bf 2017}, 1--54}~(2017).

\bibitem{delacretaz2020heavy}
Luca~V Delacretaz.
\newblock ``Heavy operators and hydrodynamic tails''.
\newblock \href{https://dx.doi.org/10.21468/SciPostPhys.9.3.034}{SciPost
  Physics {\bf 9}, 034}~(2020).

\bibitem{altland2021from}
Alexander Altland, Dmitry Bagrets, Pranjal Nayak, Julian Sonner, and Manuel
  Vielma.
\newblock ``From operator statistics to wormholes''.
\newblock \href{https://dx.doi.org/10.1103/PhysRevResearch.3.033259}{Phys. Rev.
  Res. {\bf 3}, 033259}~(2021).

\bibitem{yoshimura2023operator}
Takato Yoshimura, Samuel~J Garratt, and J~T Chalker.
\newblock ``Operator dynamics in floquet many-body systems''~(2023).
\newblock  \href{http://arxiv.org/abs/2312.14234}{arXiv:2312.14234}.

\bibitem{bouverot2024random}
Oscar Bouverot-Dupuis, Silvia Pappalardi, Jorge Kurchan, Anatoli Polkovnikov,
  and Laura Foini.
\newblock ``Random matrix universality in dynamical correlation functions at
  late times''~(2024).
\newblock  \href{http://arxiv.org/abs/2407.12103}{arXiv:2407.12103}.

\end{thebibliography}

\end{document}